\documentclass[a4paper,10pt,twoside]{cpc-hepnp}
\usepackage{amsmath, amssymb, graphicx, feynmp, adjustbox, ifthen}
\usepackage[colorlinks=true, citecolor=blue, urlcolor=blue, linkcolor=blue, breaklinks=true, pdfpagelabels=false]{hyperref}
\usepackage{subfigure}
\usepackage{float}
\usepackage{xcolor}
\usepackage{slashed}
\usepackage{multirow}
\usepackage{booktabs}
\usepackage{multicol}
\usepackage{chngcntr}

\makeatletter\g@addto@macro\bfseries{\boldmath}\makeatother

\allowdisplaybreaks

\usepackage{CJKutf8}
\newcommand{\zw}[1]{\begin{CJK}{UTF8}{gbsn}{#1}\end{CJK}}

\newcommand{\FDF}{(\varphi^\dagger i\!\!\overleftrightarrow{D}_\mu\varphi)}
\newcommand{\FDFI}{(\varphi^\dagger i\!\!\overleftrightarrow{D}^I_\mu\varphi)}

\newcommand{\ReIm}{{}_{\Re}^{[\Im]}\!}

\newcommand{\headertext}{Submitted to Chinese Physics C}
\newcommand{\preprintnumber}{}% to be put on the first page only
\newlength{\preprintwidth}
\begin{document}

% compute column width for \ruleup and \ruledown
\newlength{\mycolwidth}
\setlength{\mycolwidth}{.5\textwidth}
\addtolength{\mycolwidth}{-.5\columnsep}

% metapost diagrams

% title stuffs
\renewcommand{\preprintnumber}{\scriptsize{}}

\title{Probing the top quark flavor-changing couplings at CEPC\thanks{supported
by CNRS via the LIA FCPPL, by the CEPC theory study grant, and by IHEP
under Contract No.~Y7515540U1.}}

\author{Liaoshan Shi (\zw{石辽珊})$^{1}$\email{liaoshan.shi@cern.ch}%
    \quad Cen Zhang (\zw{张岑})$^{1,2}$\email{cenzhang@ihep.ac.cn}%
}

\maketitle

\address{%
$^1$ Institute of High Energy Physics, Chinese Academy of Sciences, Beijing
100049, China\\
$^2$ School of Physical Sciences,
University of Chinese Academy of Sciences, Beijing 100049, China\\
}
\begin{abstract}
	We propose to study the flavor properties of the top quark at the
	future Circular Electron Positron Collider (CEPC) in China.  We
	systematically consider the full set of 56 real parameters that
	characterize the flavor-changing neutral interactions of the top quark,
	which can be tested at CEPC in the single top production channel.
	Compared with the current bounds from the LEP2 data and the projected
	limits at the high-luminosity LHC, we find that CEPC could improve
	the limits of the four-fermion flavor-changing coefficients by
	one to two orders of magnitude, and would also provide similar
	sensitivity for the two-fermion flavor-changing coefficients.  Overall,
	CEPC could explore a large fraction of currently allowed parameter
	space that will not be covered by the LHC upgrade.  We show that
	the $c$-jet tagging capacity at CEPC could further improve its
	sensitivity to top-charm flavor-changing couplings.  If a signal is
	observed, the kinematic distribution as well as the $c$-jet tagging
	could be exploited to pinpoint the various flavor-changing
	couplings, providing valuable information about the flavor properties of
	the top quark.
\end{abstract}

\begin{keyword}
	top quark, flavor-changing neutral current, lepton collider
\end{keyword}

\begin{pacs}
	13.66.Bc, 14.65.Ha, 14.80.Bn
\end{pacs}

\begin{multicols}{2}
\section{Introduction}

After the discovery of the Higgs boson~\cite{Aad:2012tfa,Chatrchyan:2012xdj},
the focus of high energy physics turned to the study of its detailed properties.  While
the Higgs measurements at the Large Hadron Collider (LHC) could reach
a precision level of about 5\%$\sim$10\% \cite{Cepeda:2019klc}
(except for the
Higgs trilinear coupling), precision measurements of Higgs couplings could
benefit from the cleaner environment of a future $e^+e^-$ collider.
Among several proposals, the Circular Electron Positron Collider (CEPC) in
China~\cite{CEPC-SPPCStudyGroup:2015csa,CEPCStudyGroup:2018ghi} is proposed to
run as a Higgs factory at 240 GeV, which maximizes the
$e^+e^-\to HZ$ cross-section, producing at least a million Higgs bosons over a
period of 7 years.

Apart from the Higgs boson, the top quark could play an equally important role
in the electroweak symmetry breaking mechanism
\cite{Hill:2002ap}.
By virtue of its large mass, it is often thought of as a window to new physics.
Producing top quark pairs at a lepton collider would, however, require a minimum
center-of-mass energy of about $2m_{top}\approx$ 345 GeV, beyond the currently
planned CEPC energy.  While an energy upgrade above the $t\bar t$
threshold remains an option, an interesting question to ask is whether we
could still learn something about the top quark at an energy below the
production threshold.  One possibility, for instance, would be to study
virtual top quarks, which appear in almost all electroweak processes due to
quantum corrections
\cite{Vryonidou:2018eyv,Boselli:2018zxr,Durieux:2018ggn}.

In this work, we study a different possibility: instead of producing pairs of top
quarks on shell, single top quark can be produced in association with a light quark.
The process $e^+e^-\to t(\bar t)j$ is possible with $E_{cm}=240$ GeV.  This process is
highly suppressed by the Glashow-Iliopoulos-Maiani (GIM) mechanism \cite{Glashow:1970gm}
in the Standard Model (SM), but if physics beyond SM exists and gives rise to
the so called top quark flavor-changing neutral (FCN) interactions, this
production mode could happen via an $s$-channel $Z$ or photon, or via a contact
four-fermion FCN interaction.  The top quark FCN couplings have been searched
for at the LHC, Tevatron, LEP2 and HERA experiments
\cite{Abe:1997fz,Aaltonen:2008ac,Aaltonen:2009ef,Abazov:2011qf,Aaboud:2018nyl,Aad:2015uza,Aad:2012ij,ATLAS:2011mka,ATLAS:2011jga,CMS:2017twu,Chatrchyan:2013nwa,Chatrchyan:2012hqa,Aaboud:2018oqm,Aaboud:2018pob,Sirunyan:2017uae,Aaboud:2017mfd,Khachatryan:2016atv,Aad:2015pja,CMS:2015qhe,Aad:2014dya,CMS:2014qxa,TheATLAScollaboration:2013nia,Aaltonen:2008qr,Aad:2015gea,TheATLAScollaboration:2013vha,Aad:2012gd,Abazov:2010qk,Abazov:2007ev,Khachatryan:2016sib,Khachatryan:2015att,Sirunyan:2017kkr,Aleph:2001dzz,Abbiendi:2001wk,Heister:2002xv,Achard:2002vv,Abdallah:2003wf,DELPHI:2011ab,Abramowicz:2011tv,Chekanov:2003yt,H1,Aaron:2009vv,Aktas:2003yd}.
Currently,
the best constraints of the two-fermion FCN couplings come from the LHC
\cite{Aaboud:2018nyl,Aad:2015gea,Khachatryan:2016sib,Khachatryan:2015att,Aaboud:2018pob},
while
the four-fermion contact interactions have received much less attention,
even
though they are indispensable for a complete description of FCN couplings,
and are also motivated by the explicit models beyond SM
\cite{Durieux:2014xla,Iltan:2002re,Frank:2006ku,Han:2011xd}.
Interestingly, it was shown that
the best sensitivity for the $eetq$ contact interactions is still given by
the LEP2 experiments, despite its much lower integrated luminosity \cite{Durieux:2014xla,Cerri:2018ypt}. LHC and LEP2 thus
provide complementary constraints of the theory space spanned by the two types
of FCN interactions.  This immediately implies that a future $e^+e^-$ collider could further improve our knowledge of the top quark flavor properties.
The goal of this paper is to study the prospects of top FCN couplings at CEPC, to
demonstrate that a similar complementarity is expected between CEPC and
the high-luminosity LHC (HL-LHC), and to provide input for CEPC
experiments.  Similar prospects have been provided previously for TESLA, FCC-ee,
and CLIC \cite{AguilarSaavedra:2001ab,Khanpour:2014xla,deBlas:2018mhx},
but only the CLIC report \cite{deBlas:2018mhx} has considered the four-fermion
interactions.

The paper is organized as follows. In Section~\ref{sec:theory}, we describe the
theory background with focus on the two-fermion FCN and four-fermion $eetq$
FCN interactions, and their different sensitivities at a hadron collider and a
$e^+e^-$ collider. In Section~\ref{sec:simulation}, we give the details of our
simulation and our analysis strategy. In Section~\ref{sec:results}, we present our
results and discuss possible improvements. Section~\ref{sec:conclusion} is devoted to our
conclusion. Some additional results can be found in Appendix~\ref{sec:ad}.

\section{Flavor changing effective operators}
\label{sec:theory}

FCN interaction of the top quark is highly suppressed by the GIM mechanism.
The branching ratios for two-body top FCN decays in SM are of the order of
$10^{-12}$--$10^{-15}$ \cite{Eilam:1990zc,Mele:1998ag,AguilarSaavedra:2002ns}.
Any hint for such processes would thus
immediately point to physics beyond SM. 
A wide variety of limits have been set on these couplings.
For example, flavor changing decay modes $t\to qZ$ and
$t\to q\gamma$ were searched for at the Tevatron
by CDF \cite{Abe:1997fz, Aaltonen:2008ac, Aaltonen:2009ef} and D0
\cite{Abazov:2011qf},
and at the LHC by ATLAS \cite{Aaboud:2018nyl,Aad:2015uza,Aad:2012ij,
ATLAS:2011mka, ATLAS:2011jga}
and CMS \cite{CMS:2017twu,Chatrchyan:2013nwa, Chatrchyan:2012hqa}.
At the LHC, $t\to qH$ was also searched for
\cite{Aaboud:2018oqm,Aaboud:2018pob,Sirunyan:2017uae,Aaboud:2017mfd,Khachatryan:2016atv,Aad:2015pja,CMS:2015qhe,Aad:2014dya,CMS:2014qxa,TheATLAScollaboration:2013nia}.
Direct top production, $pp\to t$, was considered at the Tevatron by
CDF~\cite{Aaltonen:2008qr} and at the LHC by ATLAS
\cite{Aad:2015gea,TheATLAScollaboration:2013vha, Aad:2012gd}, while a similar
production with an additional jet in the final state was considered by
D0~\cite{Abazov:2010qk, Abazov:2007ev} and CMS~\cite{Khachatryan:2016sib}.
Single top production in association with a photon and a $Z$ were searched for
by CMS \cite{Khachatryan:2015att} and ATLAS \cite{Sirunyan:2017kkr}.  At LEP2,
$e^+e^-\to tj$ was investigated by all four collaborations
\cite{Aleph:2001dzz,Abbiendi:2001wk, Heister:2002xv, Achard:2002vv,
Abdallah:2003wf,DELPHI:2011ab}, while at HERA, the single-top $e^-p\to e^-t$
production was considered by ZEUS~\cite{Abramowicz:2011tv,
Chekanov:2003yt} and H1~\cite{H1, Aaron:2009vv, Aktas:2003yd}.  The most
constraining limits were recently collected and summarized in Table~33 of
Ref.~\cite{Cerri:2018ypt}.  The sensitivities in terms of the two-body branching
ratios are roughly of the order $10^{-4}$ to $10^{-3}$, approaching the expected
values from typical new physics models \cite{AguilarSaavedra:2004wm}.

A complete and systematic description of the top quark FCN couplings based on the
Standard Model Effective Field Theory (SMEFT) \cite{Weinberg:1978kz,
Leung:1984ni, Buchmuller:1985jz} was discussed and documented
in the LHC TOP Working Group note \cite{AguilarSaavedra:2018nen}.
The idea is that starting from the Warsaw basis operators
\cite{Grzadkowski:2010es}, one defines linear combinations of Wilson
coefficients that give independent contributions in a given measurement.  For the
$e^+e^-\to tj$ process, the relevant basis operators are the
following two-fermion operators:
\begin{flalign}
	&O_{\varphi q}^{1(ij)} = \FDF \left( \bar q_i \gamma^\mu q_j \right),
	\\
	&O_{\varphi q}^{3(ij)} = \FDFI \left( \bar q_i \gamma^\mu \tau^I q_j \right),
	\\
	&O_{\varphi u}^{(ij)} = \FDF \left( \bar u_i \gamma^\mu u_j \right),
	\\
	&O_{uW}^{(ij)}=\left( \bar q_i\sigma^{\mu\nu}\tau^I u_j \right)\tilde \varphi W_{\mu\nu}^I,
	\\
	&O_{uB}^{(ij)}=\left( \bar q_i\sigma^{\mu\nu} u_j \right)\tilde \varphi B_{\mu\nu},
\end{flalign}
where $\varphi$ is the Higgs doublet, $\tilde\varphi=i\sigma_2\varphi$,
$\tau^I$ is the Pauli matrix, $B_{\mu\nu}$ and $W_{\mu\nu}^I$ are the
$U(1)_Y$ and $SU(2)_L$ gauge field strength tensors,
$\FDF\equiv i\varphi^\dagger\left( D_\mu-\overleftarrow{D}_\mu \right)\varphi$,
and
$\FDF\equiv i\varphi^\dagger\left( \tau^ID_\mu-\overleftarrow{D}_\mu \tau^I\right)\varphi$.
The following four-fermion basis operators are also relevant:
\begin{flalign}
	&O_{lq}^{1(ijkl)}=\left( \bar l_i \gamma_\mu l_j \right)
			\left( \bar q_k \gamma^\mu q_l \right),
	\\
	&O_{lq}^{3(ijkl)}=\left( \bar l_i \gamma_\mu \tau^I l_j \right)
			\left( \bar q_k \gamma^\mu \tau^I q_l \right),
	\\
	&O_{lu}^{(ijkl)}=\left( \bar l_i \gamma_\mu l_j \right)
			\left( \bar u_k \gamma^\mu u_l \right),
	\\
	&O_{eq}^{(ijkl)}=\left( \bar e_i \gamma_\mu e_j \right)
			\left( \bar q_k \gamma^\mu q_l \right),
	\\
	&O_{eu}^{(ijkl)}=\left( \bar e_i \gamma_\mu e_j \right)
			\left( \bar u_k \gamma^\mu u_l \right),
	\\
	&O_{lequ}^{1(ijkl)}=\left( \bar l_i e_j \right)\varepsilon
			\left( \bar q_k u_l \right),
	\\
	&O_{lequ}^{3(ijkl)}=\left( \bar l_i \sigma_{\mu\nu} e_j \right)\varepsilon
			\left( \bar q_k \sigma^{\mu\nu} u_l \right),
\end{flalign}
where $i,j,k,l$ are flavor indices.
For the four-fermion operators,
only the $i=j=1$ components are relevant for the $e^+e^-\to t(\bar t)j$
process.
Other operators such as 
$O_{u\varphi}^{(ij)}\equiv \left( \varphi^\dagger\varphi \right)\left( \bar q_iu_j\tilde\varphi \right)$
and $O_{uG}^{(ij)} \equiv \left( \bar q_i\sigma^{\mu\nu} T^a u_j
\right)\tilde \varphi G_{\mu\nu}^a$
could lead to FCN couplings $tqH$ and $tqg$, but they cannot be probed
in the single top channel.
The following linear combinations of Wilson coefficients can be defined
as independent degrees of freedom that enter this process:

Two-fermion degrees of freedom:
\begin{flalign}
	& c_{\varphi q}^{-[I](3+a)}\equiv
	\ReIm \left\{ C_{\varphi q}^{1(3a)} - C_{\varphi q}^{3(3a)} \right\},
	\\
	& c_{\varphi u}^{[I](3+a)}\equiv
	\ReIm \left\{ C_{\varphi u}^{1(3a)} \right\},
	\\
	& c_{uA}^{[I](3a)}\equiv
	\left\{ c_WC_{uB}^{(3a)}+s_WC_{uW}^{(3a)} \right\},
	\\
	& c_{uA}^{[I](a3)}\equiv
	\left\{ c_WC_{uB}^{(a3)}+s_WC_{uW}^{(a3)} \right\},
	\\
	& c_{uZ}^{[I](3a)}\equiv
	\left\{ -s_WC_{uB}^{(3a)}+c_WC_{uW}^{(3a)} \right\},
	\\
	& c_{uZ}^{[I](a3)}\equiv
	\left\{ -s_WC_{uB}^{(a3)}+c_WC_{uW}^{(a3)} \right\}.
\end{flalign}

Four-fermion $eetq$ degrees of freedom:
\begin{flalign}
	&c_{lq}^{-[I](1,3+a)}\equiv
	\ReIm\left\{ C_{lq}^{1(113a)}-C_{lq}^{3(113a)} \right\},
	\\
	&c_{eq}^{[I](1,3+a)}\equiv
	\ReIm\left\{ C_{eq}^{(113a)} \right\},
	\\
	&c_{lu}^{[I](1,3+a)}\equiv
	\ReIm\left\{ C_{lu}^{(113a)} \right\},
	\\
	&c_{eu}^{[I](1,3+a)}\equiv
	\ReIm\left\{ C_{eu}^{(113a)} \right\},
	\\
	&c_{lequ}^{S[I](1,3a)}\equiv
	\ReIm\left\{ C_{lequ}^{1(113a)} \right\},
	\\
	&c_{lequ}^{S[I](1,a3)}\equiv
	\ReIm\left\{ C_{lequ}^{1(11a3)} \right\},
	\\
	&c_{lequ}^{T[I](1,3a)}\equiv
	\ReIm\left\{ C_{lequ}^{3(113a)} \right\},
	\\
	&c_{lequ}^{T[I](1,a3)}\equiv
	\ReIm\left\{ C_{lequ}^{3(11a3)} \right\},
\end{flalign}
where quark generation indices ($a=1,2$) and lepton generation indices are enclosed in
parentheses. 
An $I$ in the superscript represents the imaginary part of the coefficient,
denoted by $\Im$ on the right hand side, while
without $I$ only the real part is taken, represented by $\Re$
on the right hand side.
In total, one collects the following 28 real and independent degrees of freedom
for each $a$ (and thus 56 in total):
\begin{flalign}
\footnotesize
\setlength{\arraycolsep}{2.5pt}
\begin{array}{lllllll}
	c_{\varphi q}^{-(3+a)} & c_{uZ}^{(a3)} & c_{uA}^{(a3)} & 
	c_{lq}^{-(1,3+a)} & c_{eq}^{(1,3+a)} &
	c_{lequ}^{S(1,a3)} & c_{lequ}^{T(1,a3)} 
	\\
	c_{\varphi u}^{(3+a)} & c_{uZ}^{(3a)} & c_{uA}^{(3a)} & 
	c_{lu}^{(1,3+a)} & c_{eu}^{(1,3+a)} &
	c_{lequ}^{S(1,3a)} & c_{lequ}^{T(1,3a)} 
        \\
	c_{\varphi q}^{-I(3+a)} & c_{uZ}^{I(a3)} & c_{uA}^{I(a3)} & 
	c_{lq}^{-I(1,3+a)} & c_{eq}^{I(1,3+a)} &
	c_{lequ}^{SI(1,a3)} & c_{lequ}^{TI(1,a3)} 
        \\
	c_{\varphi u}^{I(3+a)} & c_{uZ}^{I(3a)} & c_{uA}^{I(3a)} & 
	c_{lu}^{I(1,3+a)} & c_{eu}^{I(1,3+a)} &
	c_{lequ}^{SI(1,3a)} & c_{lequ}^{TI(1,3a)} 
\label{eq:28}
\end{array}
\end{flalign}

Among the
seven columns, the first three come from the two-fermion operators.
$c^-_{\varphi q}$ and $c_{\varphi u}$ give rise to $tqZ$ coupling with a vector-like
Lorentz structure, while $c_{uA}$ and $c_{uZ}$ give rise to the $tq\gamma$ and $tqZ$ dipole
interactions.  The last four come from the $eetq$ four-fermion operators.
$c^-_{lq}$, $c_{lu}$, $c_{eq}$, and $c_{eu}$ coefficients give rise to interactions between
two vector currents, while $c^S_{lequ}$ and $c^T_{lequ}$ to interactions
between two scalar and two tensor currents, respectively.  
We note that the first two rows are CP-even while the last two rows are CP-odd.
The first and the third rows involve a left-handed light quark, while the second
and the fourth rows involve a right-handed light quark. 
The interference between coefficients from different rows in the limit of massless
quarks vanishes for this reason.  Furthermore, the signatures of the degrees of freedom in the first row
are identical to those in the third row, and similarly the second row is
identical to the fourth row.  This is due to the absence of an SM amplitude that
interferes with the FCN coefficients, which leads to cross-sections that are invariant
under a change of phase:
$c_i+c^I_ii\to e^{i\delta} (c_i+c^I_ii)$.
It is therefore sufficient to focus on the degrees of freedom in the first two
rows, and in the rest of the paper we will refer to them simply as
coefficients.  We also note that the $e^+e^-\to tj$ signal of the coefficients
from the first two rows are similar, up to a $\theta\to \pi-\theta$
transformation of the scattering angle in the $tj$ production.  The decay of the top quark, however,
breaks this similarity.  This is because the two coefficients produce
left-handed and right-handed top quarks respectively, while the lepton momentum
from the top decay is correlated with the top helicity.  This leads to a difference
in signal efficiencies between the first two rows.

Two-fermion FCN interactions in the first three columns are considered in
almost all experimental searches.  Four-fermion FCN interactions, on the other
hand, have unduly been neglected. They were proposed in
Ref.~\cite{BarShalom:1999iy}, and searched for at LEP2 by the L3 and
DELPHI collaborations \cite{Achard:2002vv,DELPHI:2011ab}, but the three-body
decays through four-fermion FCN interactions have never been searched for at the
Tevatron or LHC, except for the lepton-flavor violating case.  As for the prospects at
future $e^+e^-$ colliders, four-fermion couplings were also neglected in
the studies of single top at TESLA and FCC-ee
\cite{AguilarSaavedra:2001ab,Khanpour:2014xla}, although the recent CLIC yellow
report has included them \cite{deBlas:2018mhx}.  However, the four-fermion
operators are indispensable for a complete characterization of the top quark
flavor properties.  They could arise, for example, in the presence of a heavy
mediator coupling to one top quark and one light quark, or in the cases where
the equation of motion (EOM) is used to remove redundant two-fermion operators in
terms of the basis operators.  Their existence also guarantees the correctness
of the effective description when particles go off-shell or in loops, see
\cite{Durieux:2014xla} for a detailed discussion.  The three-body decay $t\to
cf\bar f$ was calculated in several explicit models
\cite{Iltan:2002re,Frank:2006ku,Han:2011xd}, giving a further motivation for
considering the $tcll$ contact operators.  Ref.~\cite{Chala:2018agk}
recasted the LHC constraints of $t\to qZ$ to provide bounds.  Finally, the
lepton-and-quark-flavor violating top decay through contact interactions was
studied in \cite{Davidson:2015zza}, and recently searched for by the ATLAS
collaboration \cite{ATLAS:2018avw}.

An interesting fact about the $eetq$ four-fermion FCN interaction is that the most
stringent limits are still coming from the LEP2 experiments.  In
Ref.~\cite{Cerri:2018ypt}, a global analysis based on the current bounds
was performed within the SMEFT framework.  The result clearly showed that the
LHC is more sensitive to the two-fermion operator coefficients, while LEP2
is more sensitive to the four-fermion ones.  Hence, their
results are currently complementary in the full parameter space, as
demonstrated in Figure~59 in Section 8.1 of Ref.~\cite{Cerri:2018ypt}.  The
complementarity persists even with HL-LHC (see Figure~59 right of Ref.~\cite{Cerri:2018ypt}), despite
an order of magnitude difference between the LEP2 and HL-LHC luminosities.
Clearly, this implies that an $e^+e^-$ collider with higher luminosity could
continue to provide valuable information about the top FCN interactions, and
explore the parameter space which will not be covered by the HL-LHC.

The difference in sensitivities between the two types of colliders can be
understood as follows.  The two-fermion operators can be searched for at
the LHC by the flavor-changing decay of the top quark, but the same decay
through a four-fermion operator is a three-body decay, and will be suppressed by
an additional phase space factor.  As an illustration, the decay rates of $t\to
ce^+e^-$ through $c_{\varphi u}$, $c_{uZ}$ and $c_{eu}$ are $8.1\times
10^{-5}$, $2.4\times10^{-4}$ GeV and $3.2\times10^{-6}$ GeV, respectively, for
$c/\Lambda^2=1$ TeV$^{-2}$. Furthermore, the $e^+e^-$ mass spectrum is a
continuum, and thus the best sensitivity requires a dedicated search
without a mass window cut (see discussions in Refs.~\cite{Durieux:2014xla,Chala:2018agk}).
Searching for four-fermion operators in single top channels at a hadron
collider suffers from the same phase-space suppression.  The situation in
an $e^+e^-$ collider is, however, different.  The two-fermion operators can be
searched for through single top $e^+e^-\to Z^*/\gamma^* \to tj$ (or through top
decay if the center-of-mass energy allows for top quark pair production, though
typically the former has a better sensitivity \cite{AguilarSaavedra:2001ab}).
In the case of a four-fermion operator case, instead of a suppression effect, the production
rate is actually enhanced due to the fact that there is one less propagator than in the two-fermion
case.  As an illustration, the single top production cross-section at
$E_{cm}=240$ GeV for $c_{\varphi u}$, $c_{uZ}$ and $c_{eu}$ are 0.0018 pb,
0.020 pb and 0.12 pb, respectively, for $c/\Lambda^2=1$ TeV$^{-2}$, and this
enhancement effect increases with energy.  The comparison of the two cases is
illustrated in Figure~\ref{fig:2vs4}. Another advantage of a lepton
collider is that one can reconstruct the missing momentum.  This is not
relevant for the problem at hand, but could be important for setting
bounds on four-fermion operator with neutrinos, see
Ref.~\cite{Alcaide:2019pnf}.

\begin{center}
\includegraphics[width=0.6\linewidth]{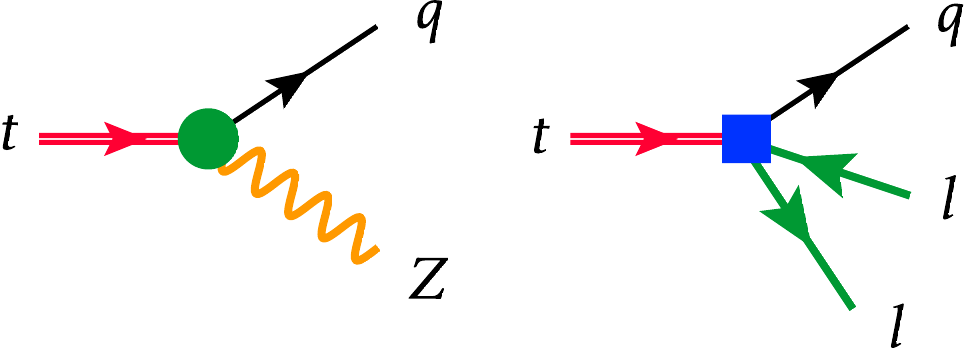}
\\[10pt]
\includegraphics[width=0.6\linewidth]{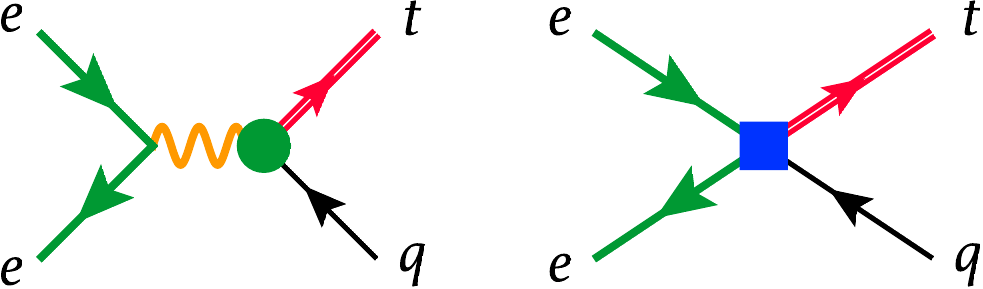}
\\[10pt]
\figcaption{\label{fig:2vs4}
(top) The flavor-changing decay at the LHC.  The four-fermion operator contribution
is suppressed by an additional phase space factor compared with the two-fermion
contribution.  (bottom) The flavor-changing single top at a $e^+e^-$ collider.  The
four-fermion operator contribution is enhanced due to one less $s$-channel
propagator than in the two-fermion case.  Green dots and blue
squares represent two- and four-fermion operator insertions.} 
\end{center}

\section{Simulation}
\label{sec:simulation}

To study the prospects of top FCN couplings, we consider the scenario of CEPC
running with a center-of-mass energy $E_{cm}=240$ GeV and an integrated
luminosity of 5.6 ab$^{-1}$.  We simulate the signal and background at leading order with
parton shower, by using {\sc MadGraph5\_aMC@NLO} \cite{Alwall:2014hca} and {\sc
Pythia8} \cite{Sjostrand:2006za,Sjostrand:2007gs}.  The signal is generated
with the {\sc UFO} model \cite{Alloul:2013bka,Degrande:2011ua}, {\sc dim6top},
which follows the LHC TopWG EFT recommendation \cite{AguilarSaavedra:2018nen}
and is available at \url{https://feynrules.irmp.ucl.ac.be/wiki/dim6top}.
The detector level simulation is performed with {\sc Delphes} with the default CEPC
card \cite{deFavereau:2013fsa}.  Jets are reconstructed using the {\sc FastJet}
package \cite{Cacciari:2011ma} with the anti-$k_t$ algorithm
\cite{Cacciari:2008gp} with a radius parameter of 0.5.  Automatic calculation for
QCD corrections of processes involving only two-fermion FCN operators were
developed in Ref.~\cite{Degrande:2014tta} (see also
Refs.~\cite{Liu:2005dp,Gao:2009rf,Zhang:2011gh,Li:2011ek,Wang:2012gp,Drobnak:2010wh,Drobnak:2010by,Zhang:2013xya,Zhang:2014rja}
where the results for the other top flavor-changing channels have been presented).  
The corrections for four-fermion operators were given in the appendix
of Ref.~\cite{Durieux:2014xla}. The sizes are below 20\%, corresponding to
less than 10\% change in the coefficients,
and therefore we neglect these corrections in this work.
The dominant background comes from the $W$-pair and
$Z$-pair production, and we do not expect a significant change at the next-to-leading order in QCD.

We consider the semi-leptonic top quark decays.  The signal final state
is $bjl\nu$, where $j$ is an up or charm quark jet.
The dominant background is $qq'l\nu$, with one light or charm quark jet
misidentified as a $b$-jet. A large fraction comes from the $W$ pair production
with one $W$ decaying hadronically and the other leptonically, 
while the diagrams with only one $W$ resonance decaying leptonically also make an
important contribution.  We thus take into account the full contribution from the
$e^+e^-\to Wq\bar q'$ process with $W$ decaying leptonically.
Adding all the diagrams from $e^+e^-\to l\nu q\bar q'$
does not make a sizable change to the background \cite{AguilarSaavedra:2001ab},
and so they are not taken into account.
Another source of background comes from $b\bar bll$ and $c\bar cll$, where one
of the jets is mistagged and one of the leptons is missed by the detector.
This is included in our simulation, but the contribution is subdominant.
Selected diagrams for the signal and background are shown in
Figure~\ref{fig:fdiagram}. 

\begin{center}
\includegraphics[width=.8\linewidth]{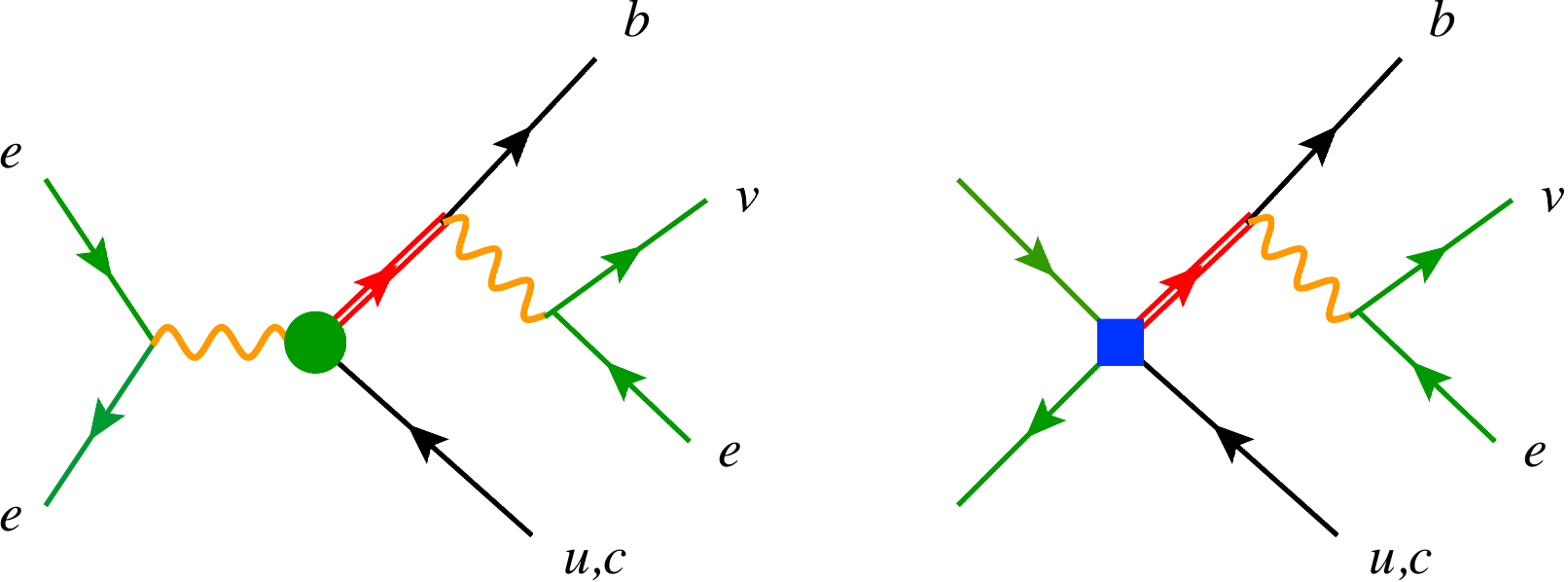}
\\[15pt]
\includegraphics[width=.7\linewidth]{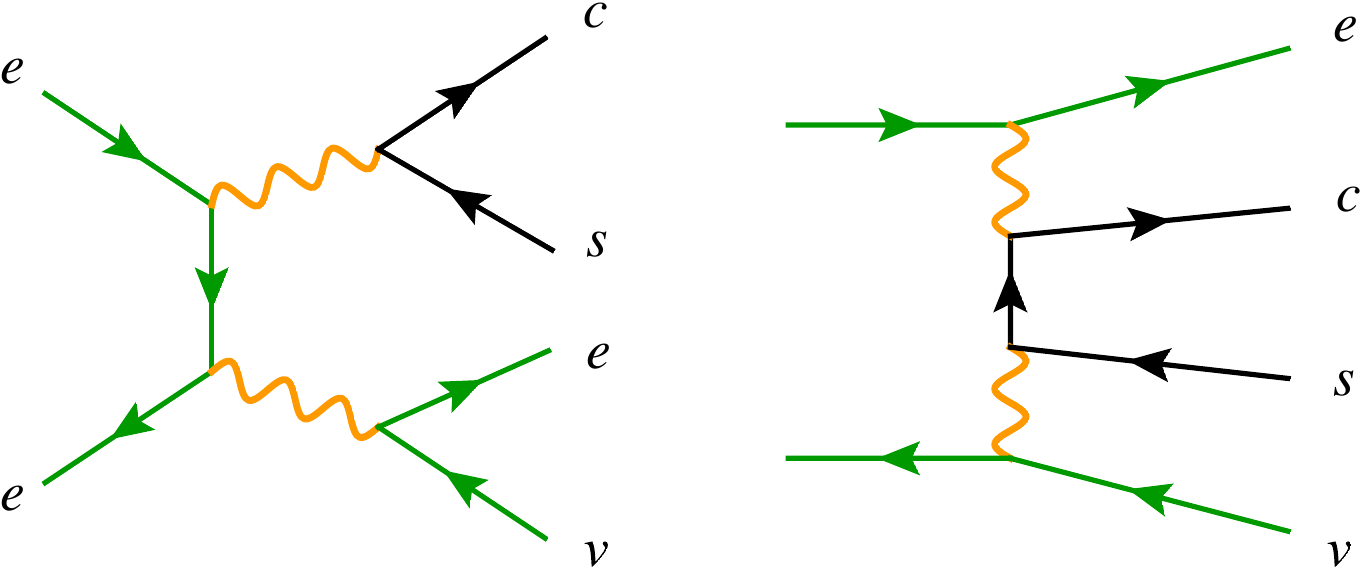}
\figcaption{\label{fig:fdiagram}
Selected Feynman diagrams for the signal (top) and background (bottom).
Green dots and blue squares represent two- and four-fermion operator insertions.
Red double lines represent top quark propagators.
}
\end{center}

Based on the expected signature of the signal process, 
we select events with exactly one charged lepton (electron or muon) and at least two jets.
The charged lepton must have $p_\text{T}>10$ GeV and $|\eta|<3.0$.
All jets are required to have $p_\text{T}>20$ GeV and $|\eta|<3.0$.
Exactly one jet should be $b$-tagged.
If more than one non-$b$-tagged jet is present, the one with the highest $p_\text{T}$ is selected
as the up or charm quark jet candidate.
We have chosen a $b$-tagging working point with 80\% efficiency for $b$-jets and a mistagging rate of 10\% (0.1\%)
from $c$-jets (light jets) \cite{Ruan:2018yrh}.
A missing energy greater than 30 GeV is also required due to the presence of a neutrino.
The $W$ boson candidate is reconstructed from the charged lepton and the missing energy.
The top quark candidate is reconstructed by combining the $W$ boson candidate with the $b$-jet.

At the parton level, we
expect the non-$b$-tagged jet from the signal to have
$E_j=\frac{s-m_{top}^2}{2\sqrt{s}}\approx 58\ \mbox{GeV}$.
For the background, if the contribution comes from the diboson production
(e.g.~Figure~\ref{fig:fdiagram} down left), we expect the dijet mass to peak at
$m_W=80.4$ GeV.  
The contribution from the non-resonant diagrams
(e.g.~Figure~\ref{fig:fdiagram} down right) cannot, however, be neglected and
gives rise to a continuum spectrum in the dijet mass distribution.
At the reconstruction level, it turns out that the energy of the non-$b$-tagged jet $E_j$,
the invariant mass of the $b$-jet and the non-$b$-tagged jet $m_{jj}$, and
the invariant mass of the top quark candidate $m_{top}$
are the most useful variables to discriminate the
signal from background. In Figure~\ref{fig:svsb}, we plot these variables at
the reconstruction level, for the background as well as for the signals from the two
typical operator coefficients, $c_{uZ}$ and $c_{eq}$, for illustration.

As our baseline analysis, we impose the following kinematic cuts at the reconstruction level
\begin{flalign}
	&E_j< 60\ \mathrm{GeV}\,,\\
	&m_{jj}>100 \ \mathrm{GeV}\,,\\
	&m_{top}<180 \ \mathrm{GeV}\,.
\end{flalign}
These cuts are motivated by Figure~\ref{fig:svsb}. 
The expected number of background events after event selection is about 1400
with an integrated luminosity of 5.6 ab$^{-1}$,
corresponding to a statistical uncertainty of about 2.7\%.
We assume that the systematic uncertainty will be under control below this
level. The impact of the systematic uncertainty can be easily estimated, e.g.~a $3\%$
systematic uncertainty will weaken the bound on the cross-section by a factor of
about 1.5, which corresponds to a factor of 1.2 on the value of the coefficients.
In the rest of the paper we simply ignore the systematic effects.
We will see that this simple baseline scenario already allows to
obtain reasonable sensitivities.

In the absence of any FCN signal, the 95\% confidence level (CL) upper bound
of the fiducial cross-section is $0.0134$ fb.
Alternatively, the 5$\sigma$ discovery limit of the signal cross-section,
determined by $S/\sqrt{B}=5$, is a function of the integrated luminosity
$L_\text{int}$:
\begin{flalign}
    \sigma = \frac{5\sqrt{\sigma_{B}}}{\sqrt{L_\text{int}}} 
               = \frac{2.51\ \mbox{fb}}{\sqrt{L_\text{int}/\mbox{fb}^{-1}}}
\end{flalign}

\begin{center}
\includegraphics[width=.9\linewidth]{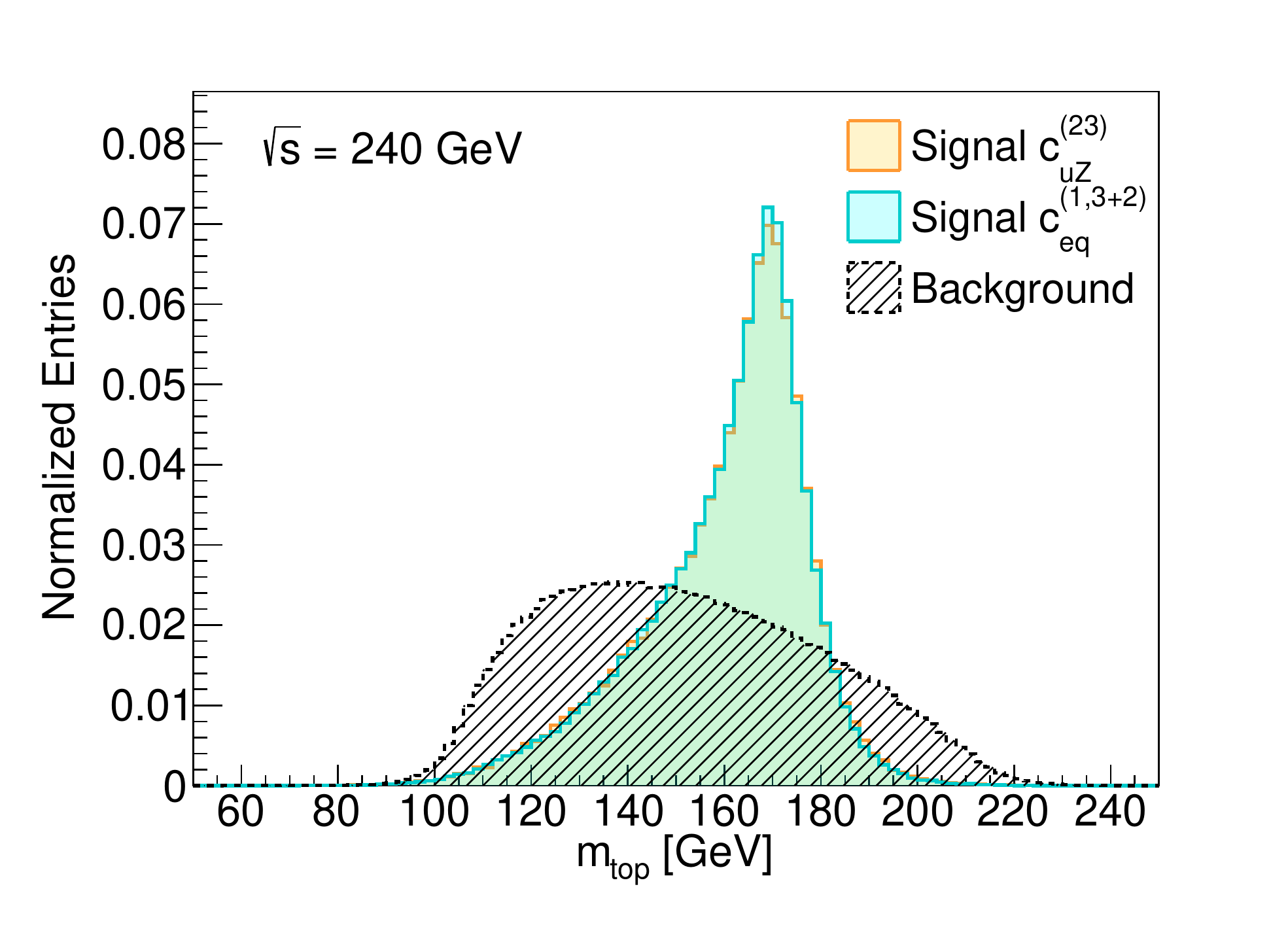}
\includegraphics[width=.9\linewidth]{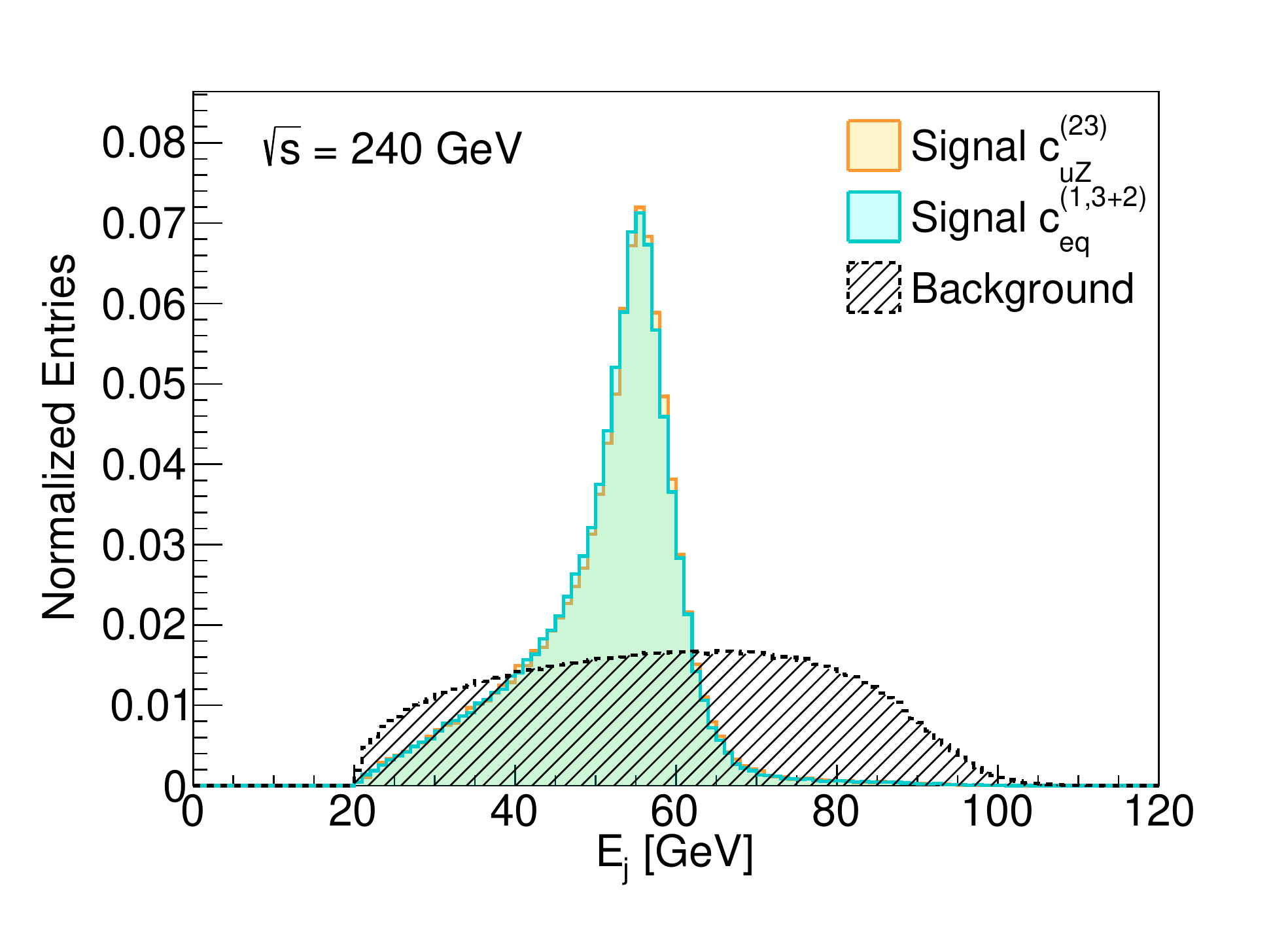}
\includegraphics[width=.9\linewidth]{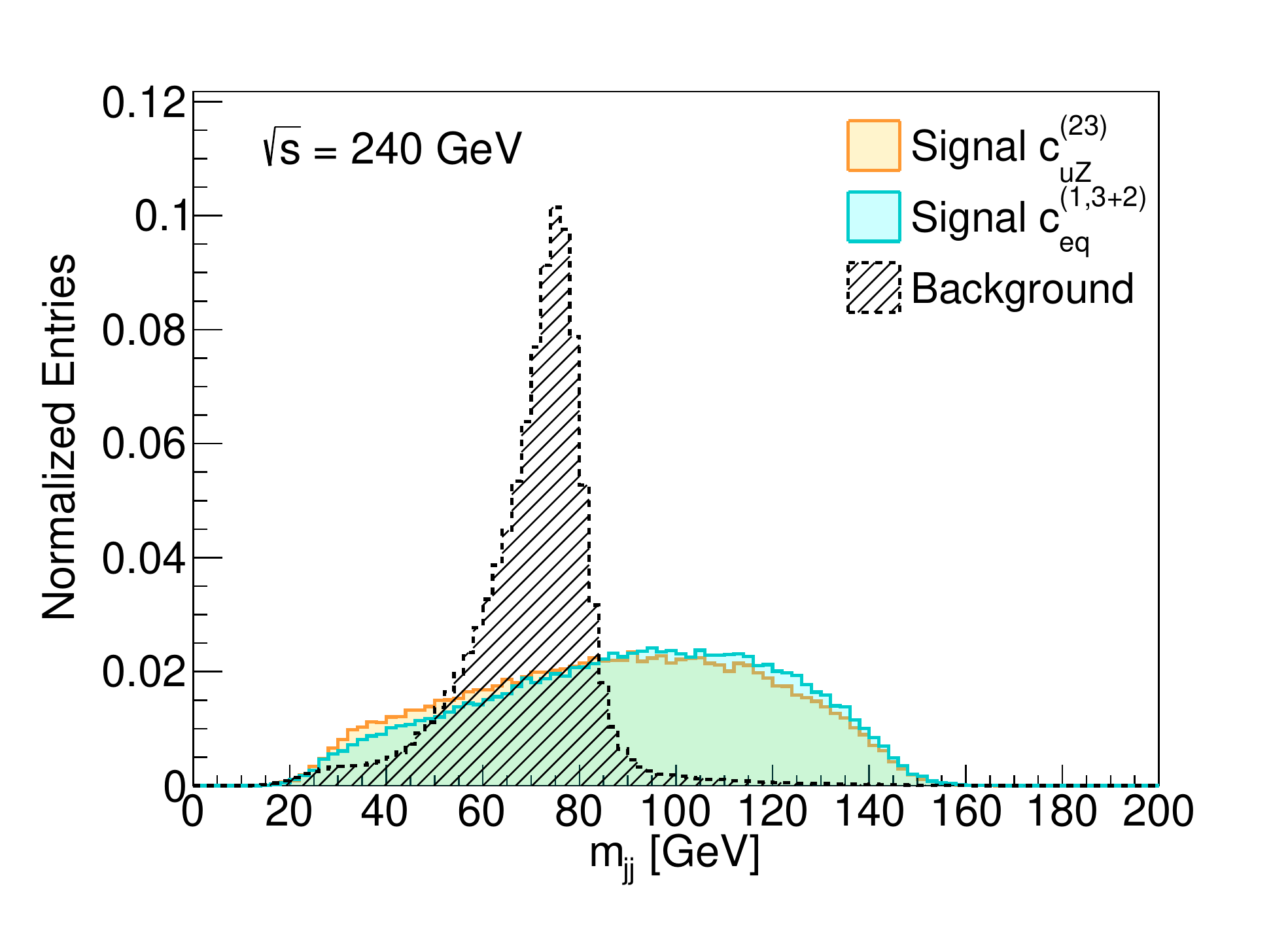}
\figcaption{\label{fig:svsb}
Signal and background at the reconstruction level. Distributions of
$m_{top}$, $E_j$, and $m_{jj}$ are shown for signals from
$c_{uZ}^{(23)}$ and $c_{eq}^{(1,3+2)}$.
}
\end{center}

The cross-section is a quadratic function of the operator coefficients.
Including the interference effects, such a function has 28 independent
terms for the 7 coefficients in each row of Eq.~(\ref{eq:28}).  These terms for the
first two rows are the same as those for the last two rows, because they only differ by
a CP phase which would never show up in the cross-section (without any possible interference
with SM).  Thus, only 56 independent terms need to be determined for the
first two rows for each $a$.  We sample the parameter space by 56 points and
simulate the fiducial cross-section for each of them.  The results are fitted
by the following form:
\begin{flalign}
	\sigma=\sum_{a=1,2}\frac{(1\ \mathrm{TeV})^4}{\Lambda^4}\left(\vec{C_1^a}\cdot\pmb{M_1^a}\cdot\vec{C_1^a}^{T}
+\vec{C_2^a}\cdot\pmb{M_2^a}\cdot\vec{C_2^a}^{T}\right)
\label{eq:mdefinition}
\end{flalign}
where $\vec{C}_{1,2}$ denote the vectors formed by the coefficients in the first and 
second rows of Eq.~(\ref{eq:28}). $a$ is the light quark generation.
$\pmb{M}_{1,2}^a$ are $7\times7$ matrices.
The above result allows to convert the upper bound and discovery limit of the
cross-section into a 56-dimensional coefficient space.

We have verified the relations between signatures from
different rows in Eq.~(\ref{eq:28}): the 1st (2nd) and the 3rd (4th) rows
always give the same signatures; the 1st (3rd) and the 2nd (4th) rows
at the production level are identical up to a $\theta\to \pi-\theta$
transformation in the production angle, but differ if the top decays.
In Appendix~\ref{sec:ad},
a comparison between the signals from $c_{uZ}^{(23)}$, $c_{uZ}^{(32)}$ and
$c_{uZ}^{I(23)}$ are shown in Figure~\ref{fig:uzcomparison}.
A comparison between the signals from $c_{eq}^{(1,3+2)}$, $c_{eu}^{(1,3+2)}$ and
$c_{eq}^{I(1,3+2)}$ are shown in Figure~\ref{fig:eqcomparison}.

\begin{center}
\begin{figure*}[ht]
\centering
\includegraphics[width=.32\linewidth]{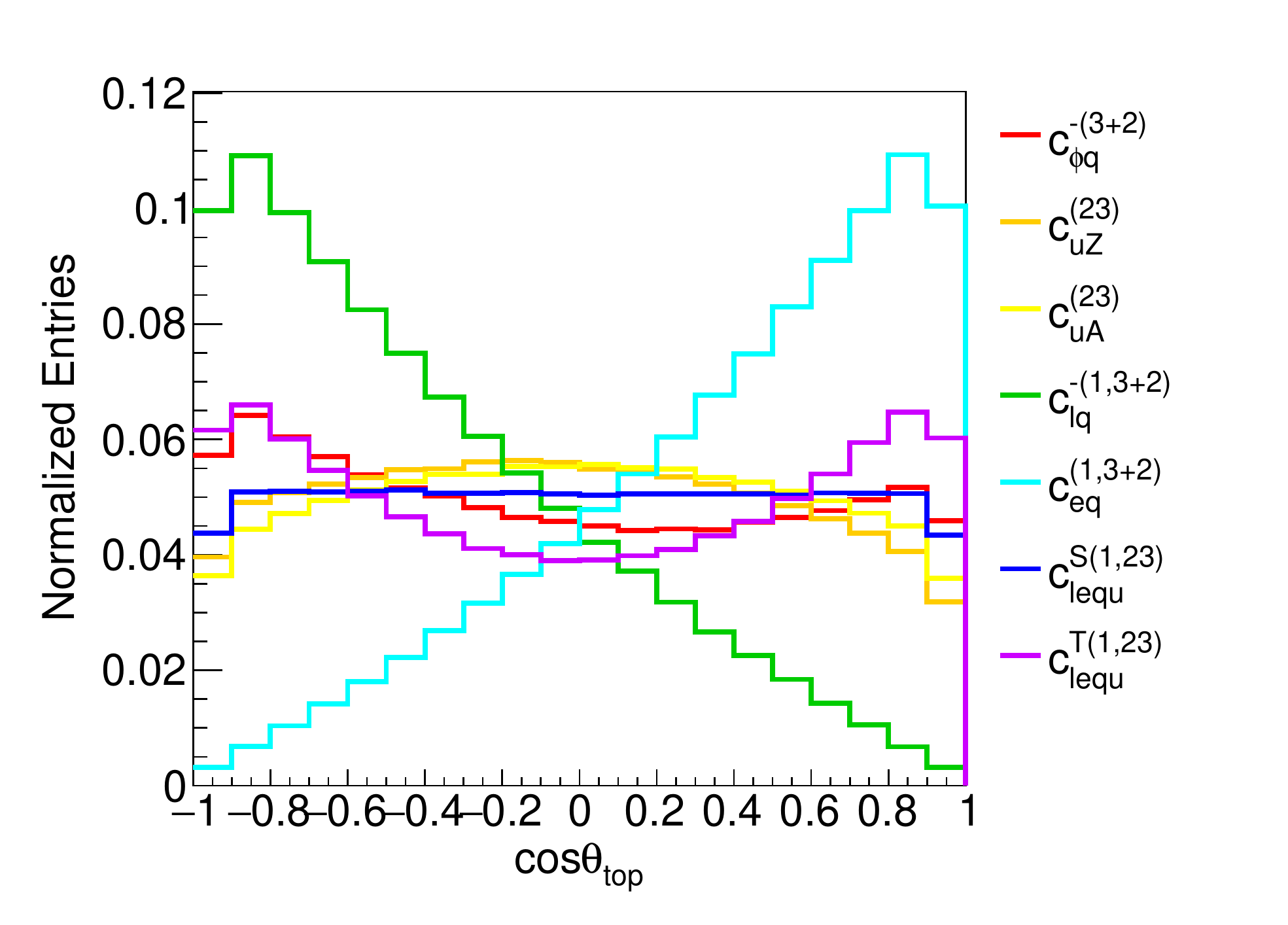}
\includegraphics[width=.32\linewidth]{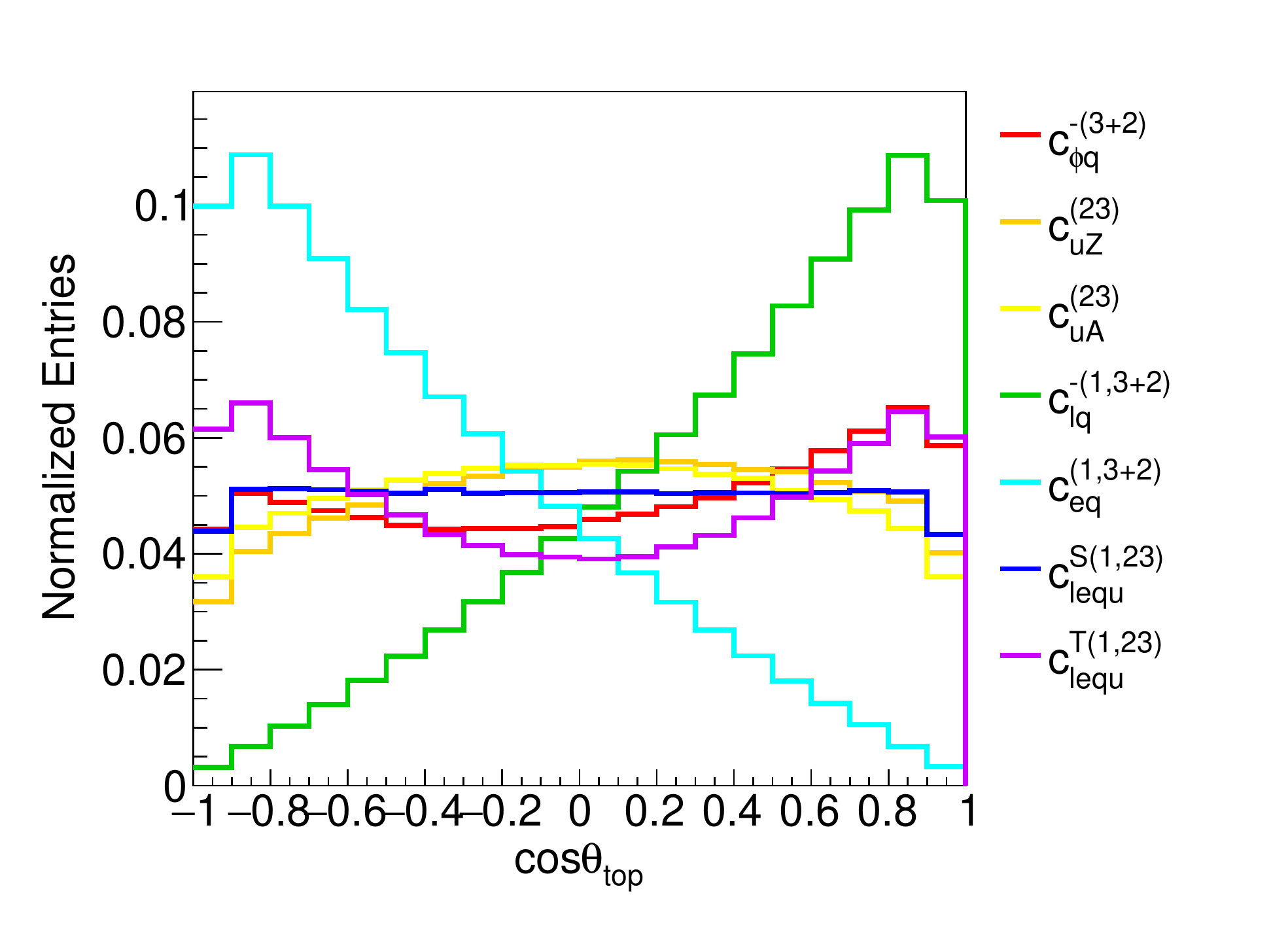}
\includegraphics[width=.32\linewidth]{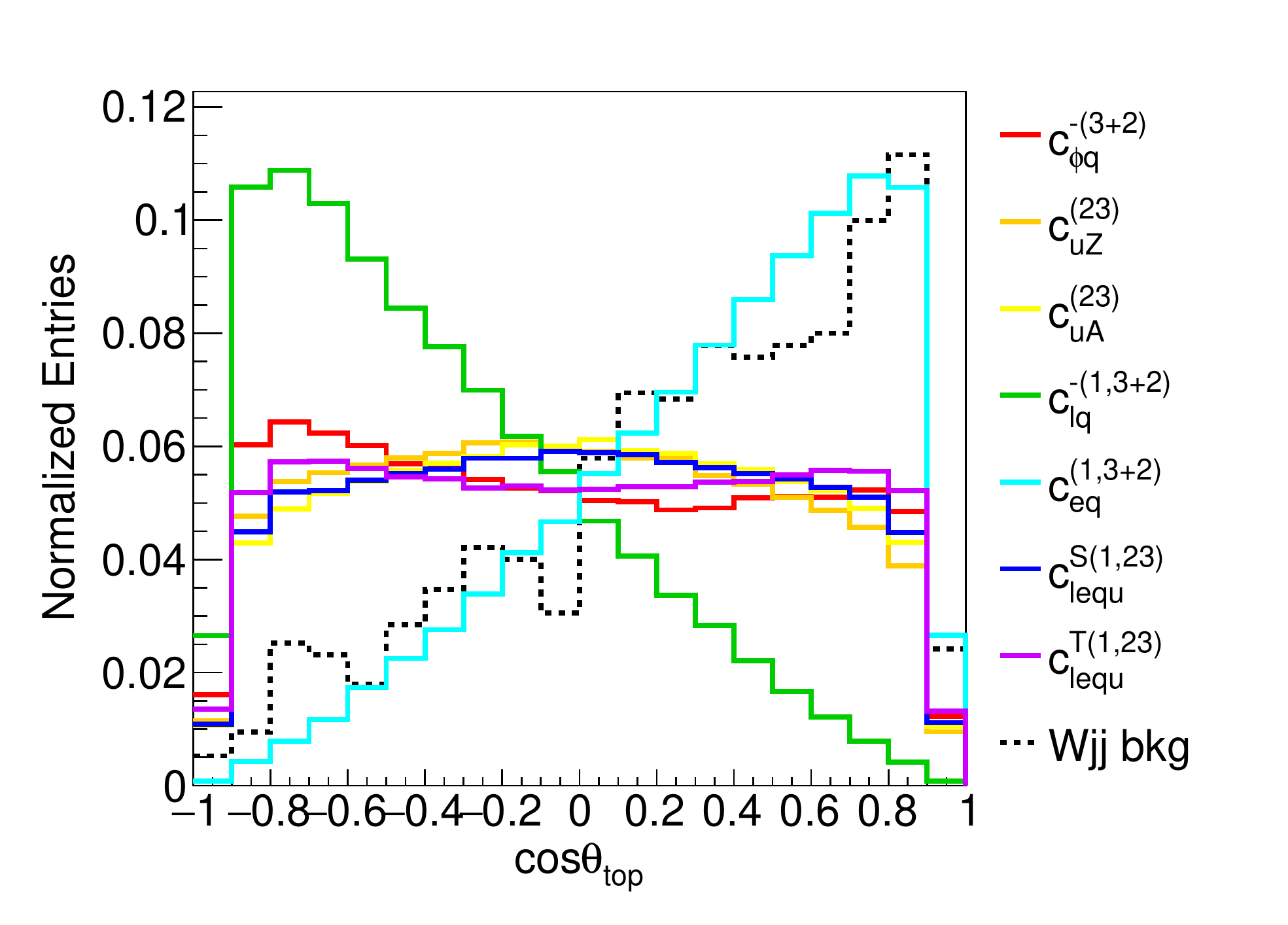}
\figcaption{\label{fig:angular}
Scattering angle from the signals of the seven coefficients in the first row
of Eq.~(\ref{eq:28}). $\theta_{top}$ is defined as the angle
between the momentum of the $e^+$ beam and $t$ or $\bar t$.
(left) parton level, for top production. (middle) parton level, for anti-top
production. (right) reconstruction level, for top production, including the
background.
}
\end{figure*}
\end{center}

Our baseline analysis could be improved by exploiting additional features of the
signal with a template fit.
One possibility is to make use of heavy flavor tagging.
The operators with $a=2$, requiring a tagged $c$-jet in the signal definition,
could largely suppress the background, as most background comes from events
with one charm and one strange quark in the final state, with the charm mistagged
as a $b$.
The clean environment of CEPC allows a precise determination of the
displaced vertices and excellent capability of $c$-jet tagging \cite{CEPCStudyGroup:2018ghi}.
We assume a working point with a 70\% tagging efficiency for $c$-jets and
20\% (12\%) mistagging rate from $b$-jets (light jets) \cite{Ruan:2018yrh}.  To constrain the coefficients
with $a=2$, we require a $c$-jet in the signal definition, while to constrain the
$a=1$ coefficients we veto the events with a $c$-jet, although the latter is not
expected to significantly change the sensitivity as most background events do
not have an extra $c$-jet except the one that fakes the $b$-jet.
Another useful information is the angular distribution of the single
top, which is determined by the specific Lorentz structure of the operator.  In
Figure~\ref{fig:angular}, we show the distribution of the top scattering angle from all 7
coefficients in the first row at the parton level and the
reconstruction level. The scattering angle $\theta$ is defined as the angle between the
momentum of the $e^+$ beam and $t$ or $\bar t$.
The distributions for the top and anti-top are related 
by $\theta\to\pi-\theta$, and this is illustrated by comparing the first two
plots in Figure~\ref{fig:angular}. Furthermore, this holds even for the reconstructed
top and anti-top candidates from the background due to the CP symmetry.
For this reason, we consider the observable $c=Q_l\times\cos\theta$,
i.e.~the lepton charge times the cosine of the scattering angle.
The discrimination power of this observable is illustrated in the right plot of
Figure~\ref{fig:angular}, at the reconstruction level.  We perform a
template fit by further dividing the signal region into 4 bins, defined as
$c\in(-1,-0.5)$, $[-0.5,0)$, $[0,0.5)$, and $[0.5,1)$.  To construct a $\chi^2$
fit, we take $\sqrt{B}$ in each bin as the experimental uncertainty.  The
smallest number of events in one bin is 24 even after requiring a $c$-jet,
and so the Gaussian distribution is a good approximation.  We simulate the
Gaussian fluctuation in all bins by generating a large number of
pseudo-measurement samples and compute the average $\chi^2$ for each point
in the coefficient space.  Our 95\% CL bound is determined by
$\left<\chi^2\right><9.49$. 

\end{multicols}%
\vspace{2mm}
%\ruleup[\mycolwidth]
%\ruledown[\mycolwidth]
\vspace{5mm}
\begin{multicols}{2}%

\section{Results}
\label{sec:results}

Following our baseline analysis, the 95\% CL limits of the individual coefficients
in the first row are given in Figure~\ref{fig:individuallimits}, where they
are compared with the current limits from LHC+LEP2 and with the HL-LHC
projection.  FCC-ee projection at the center-of-mass energy of 240 GeV is
given in Ref.~\cite{Khanpour:2014xla}, but only for the 3 two-fermion
coefficients, and we show them in the same plot.  
Note that Ref.~\cite{Chala:2018agk} suggested that the current signal region
of the $t\to ql^+l^-$ decay mode, designed for the search of $t\to qZ$ mode,
can be extended by including the ``off-shell'' region with
$\left|m_{l^+l^-}-m_Z\right|>15$ GeV.  This could lead to HL-LHC prospects that are
slightly better than Figure~\ref{fig:individuallimits} for some of the four-fermion
coefficients.
The CLIC bounds, on the other hand, are only available with higher
center-of-mass energy runs and are not shown in the plot.  For example, the
expected limits of the four-fermion coefficients from a 380 GeV run with an
integrated luminosity of 500 fb$^{-1}$, are about a factor of $3\sim4$ better
than those from CEPC, due to the higher beam energy and beam
polarization \cite{deBlas:2018mhx}.

Looking at the 3 two-fermion coefficients on the left, the limits are either
weaker than or comparable to HL-LHC.  Still, we emphasize that even in this
case the CEPC measurement provides an important consistency check with the
existing results.  The most interesting result, however, is the improvement of
the other four four-fermion coefficients.  As expected, we see that they are
1$\sim$2 orders of magnitude better than the current limits and the combination
of HL-LHC and LEP2.  Similar results are observed for the second row operators
and are displayed in Figure~\ref{fig:individuallimits2} in
Appendix~\ref{sec:ad}.  In Figure~\ref{fig:2dlimits}, we show the
two-dimensional bound of the two-fermion coefficient $c_{\varphi q}^{-(3+a)}$
and the four-fermion coefficient $c_{eq}^{(1,3+a)}$, compared with LHC, HL-LHC,
and LEP2.  Clearly, a large fraction of the currently allowed parameter space
will be probed by CEPC.  A similar plot for the operators in the second row
of Eq.~(\ref{eq:28}) is given in Figure~\ref{fig:2dlimits2} in
Appendix~\ref{sec:ad}.

In Figure~\ref{fig:discovery} we plot the discovery limits of the seven
coefficients in the first row of Eq.~(\ref{eq:28}) in terms of
$\Lambda/\sqrt{c}$, as a function of integrated luminosity.  The scale is
roughly that of new physics, assuming that the coupling is of the order of one.  The
plot shows that new physics at a few TeV leading to four-fermion FCN
interactions can be discovered already at an early stage of CEPC running.  The
improvement with luminosity is, however, less significant.
Note that the two curves corresponding to $c_{eq}^{(1,3+2)}$ and
$c_{lq}^{-(1,3+2)}$ overlap with each other.
This is because they give rise to four-fermion couplings that only
differ in the chirality of the electron fields, and thus have the same
rate in the signal region defined by our baseline analysis.
The results for the coefficients of the second row are given in
Appendix~\ref{sec:ad}, Figure~\ref{fig:discovery2},
where a similar degeneracy between $c_{eu}^{(1,3+2)}$ and $c_{lu}^{(1,3+2)}$
can be observed.

\end{multicols}%
%\vspace{2mm}
%\ruleup[\mycolwidth]
\begin{center}
%\begin{figure*}[ht]
\centering
\includegraphics[width=.8\linewidth]{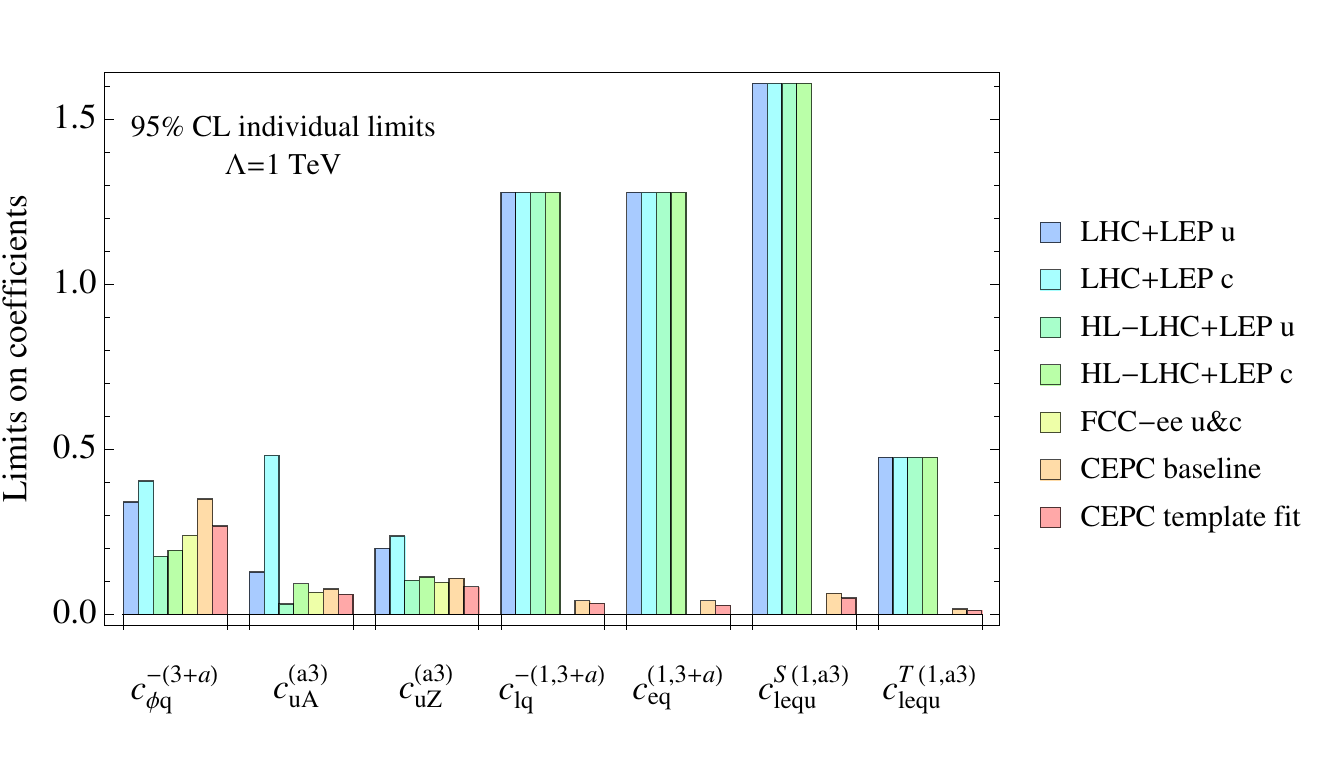}
\figcaption{\label{fig:individuallimits}
The 95\% CL limits of the individual coefficients in the first row of
Eq.~(\ref{eq:28}), as expected from CEPC, compared with the existing LHC+LEP2
bounds and the projected limits from HL-LHC+LEP2 and FCC-ee with 3 ab$^{-1}$
luminosity at 240 GeV (only for the first three coefficients), see
Refs.~\cite{Cerri:2018ypt,Khanpour:2014xla}.  The results for both generations
$a=1,2$ are displayed.  The orange column ``CEPC baseline'' is the expected
limit following our baseline analysis, which applies to both flavors (a=1,2).
The red column ``CEPC template fit'' uses $c$-jet tagging for signal
definition and only applies to $a=2$ operators.
}
%\end{figure*}
\end{center}
%\ruledown[\mycolwidth]
\vspace{1mm}
\begin{multicols}{2}%

\begin{center}
\centering
\includegraphics[width=.8\linewidth]{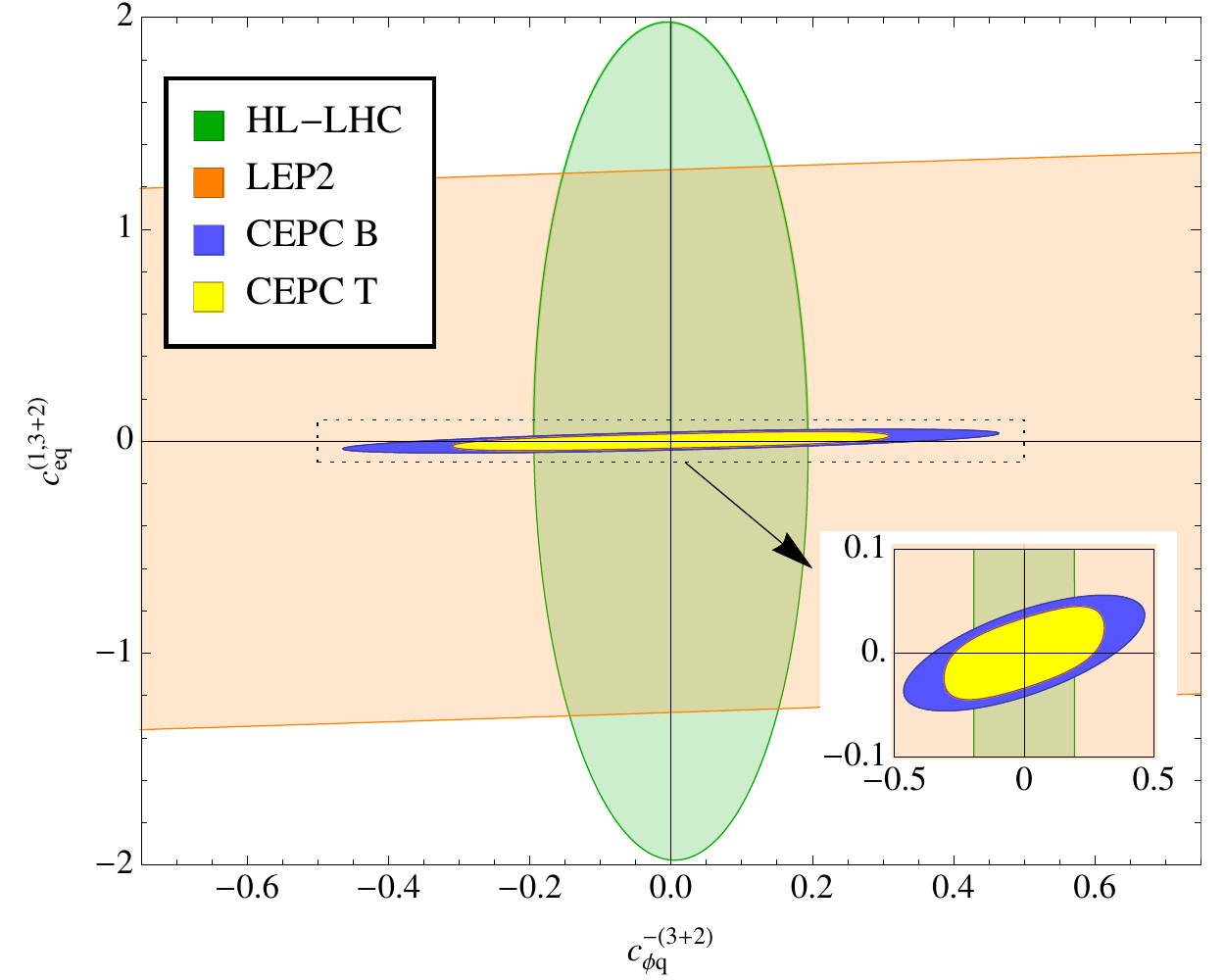}
\figcaption{\label{fig:2dlimits}
Two-dimensional bound of the two-fermion coefficient $c_{\varphi q}^{-(3+2)}$
and the four-fermion coefficient $c_{eq}^{(1,3+2)}$ at 95\% CL.
Other operators are fixed to 0. The allowed regions from HL-LHC and
LEP2 are similar to Figure~59 in Section 8.1 of Ref.~\cite{Cerri:2018ypt},
except that there all coefficients are marginalized over. The blue region
(``CEPC B'') is the bound expected from CEPC following our baseline
analysis.  The yellow region (``CEPC T'') is obtained with a template fit
approach, see the discussion in Section~\ref{sec:results}.
}
\end{center}

\begin{center}
\includegraphics[width=.95\linewidth]{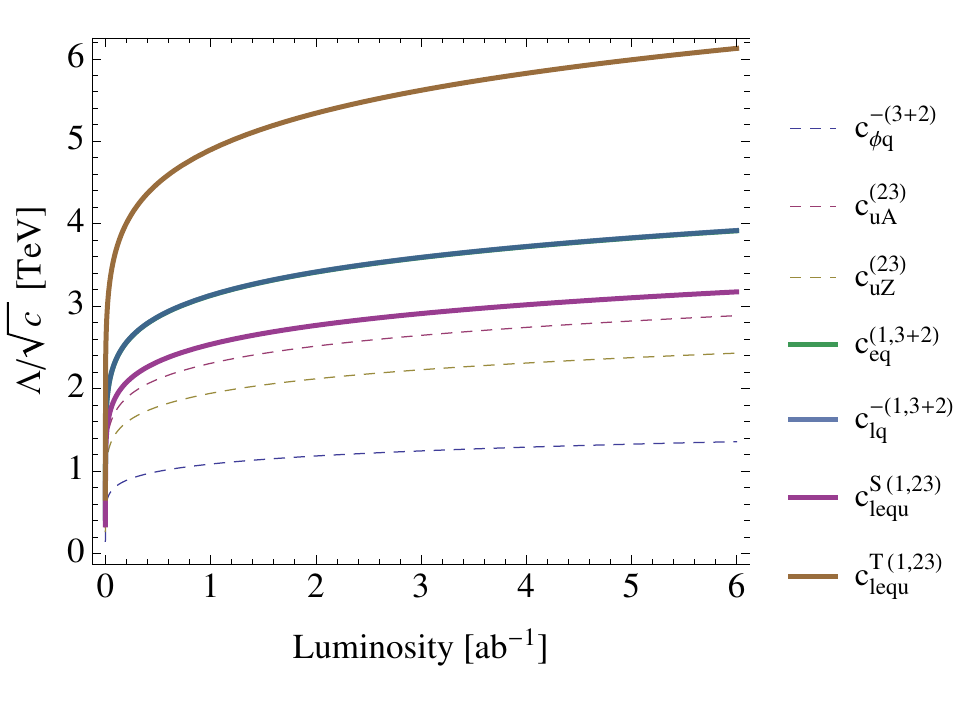}
\figcaption{\label{fig:discovery}
The five-sigma discovery limit of $\Lambda/\sqrt{c}$, which is roughly the scale
of new physics, for the coefficients in the first row of Eq.~(\ref{eq:28}), as a
function of the integrated luminosity of CEPC.
}
\end{center}

The template fit method described in the previous section leads to two-fold improvements.
First, if SM is assumed, the 95\% CL limits of the operator
coefficients for $a=2$ are improved.  This is mostly due to the $c$-tagging
requirement.  The results are shown in Figure~\ref{fig:individuallimits} (red
columns), and Figure~\ref{fig:2dlimits} (the yellow region), where the
improvements are seen clearly.  The same effects on the other four-fermion
operators are displayed and compared in Figure~\ref{fig:templatefit}.  The
second improvement is from the discrimination power between the different kinds of
signals, which comes from both the angular distribution and the $c$-tagging
information.
This is particularly important when an excess is found, in which
case we need to understand the FCN operator that leads to it.  The
baseline approach can only give the overall magnitude of the
flavor-changing effects, while the template fit helps to pin down the
actual form of the operator.  This is illustrated in
Figures~\ref{fig:discrimination}, where we consider two hypothetical scenarios,
with $c_{eq}^{(1,3+a)}=c_{lq}^{-(1,3+a)}=0.05$, and $c_{lequ}^{S(1,a3)}=0.065$,
$c_{lequ}^{T(1,a3)}=0.025$ ($\Lambda=1$ TeV). These values are
consistent with the current bounds, but are around the sensitivity expected at
CEPC.  Assuming that the other coefficients vanish, with the baseline approach
we are able to identify the overall flavor-changing effect,
but not the value of each coefficient.  The allowed region in the
two-dimensional parameter space is a ring, giving no information about the actual
form of new physics. The template fit, on the other hand, can pinpoint with
more precision the value of each coefficient.  This holds also for the $a=1$
case, even though the precision is slightly worse.  A four-fold degeneracy
shows up in the first scenario.  This is because the overall sign of the
coefficients does not have a visible effect (due to the absence of SM interference),
and the relative sign between $c_{eq}^{(1,3+a)}$ and $c_{lq}^{-(1,3+a)}$ cannot
be observed because the two operators do not interfere.  In the second case
this is reduced to a two-fold degeneracy.  This is because the interference
between $c_{lequ}^{S(1,a3)}$ and $c_{lequ}^{T(1,a3)}$ is proportional to
$\cos\theta$, so the opposite sign can be excluded by the angular
distribution. In fact, due to the shape of the background (see
Figure~\ref{fig:angular} right), the template fit has a better discrimination
power when $c_{lequ}^{S(1,a3)}$ and $c_{lequ}^{T(1,a3)}$ have opposite signs.
This effect can be seen even with the SM hypothesis, see the right plot in
Figure~\ref{fig:templatefit}.  The discrimination between $a=1$ and $a=2$ operators
is also possible with the help of $c$-tagging.  This is demonstrated in
Figure~\ref{fig:uc}, where we consider three hypothetical scenarios, with
$\left(c_{lq}^{-(1,3+1)},c_{lq}^{-(1,3+2)}\right)=(0,0.05),\ (0.05,0)$, and
$(0.35,0.35)$.  By using events with and without a $c$-jet, we can resolve the
light-quark flavor involved in the FCN coupling with some precision.  This is
unlike LHC, where one has to combine the production and decay
measurements to disentangle the two light-quark flavors in the flavor-changing
signal by using the fact that the production channel depends on the
light-quark parton distribution function.

As an additional remark, we note that a flat direction exists between the three
coefficients $c_{\varphi q}^{-(3+a)}$, $c_{lq}^{-(1,3+a)}$ and
$c_{eq}^{(1,3+a)}$, which cannot be constrained by a single run at 240 GeV.  A
second working point with larger energy would be useful to lift the degeneracy,
as the two-fermion and four-fermion contributions depend differently on energy.
All other directions can be constrained simultaneously at 240 GeV.

\begin{center}
\includegraphics[width=\linewidth]{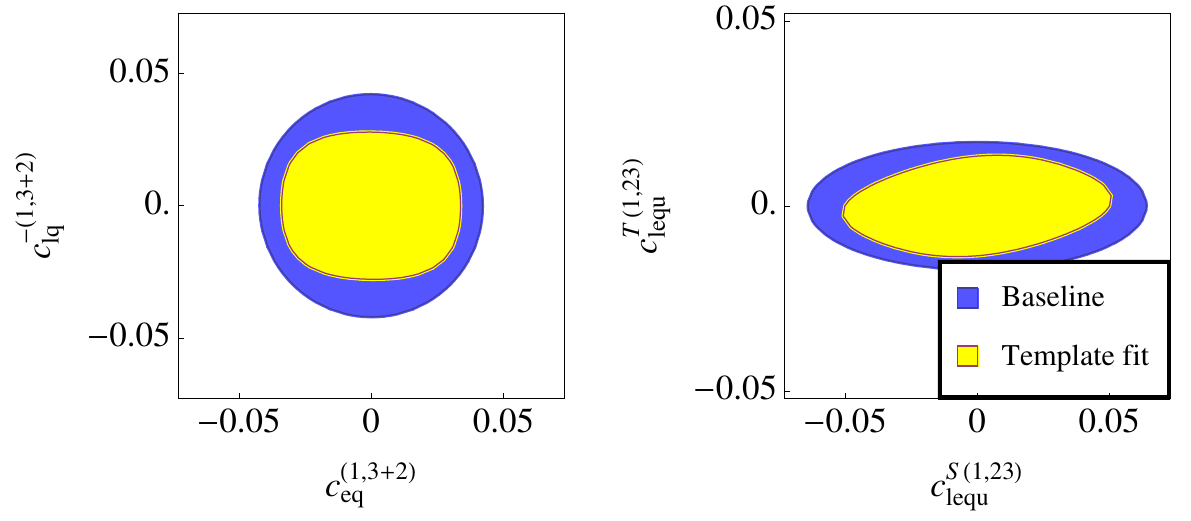}
\figcaption{\label{fig:templatefit}
Two-dimensional limits of the four-fermion coefficients, at 95\% CL,
under the SM hypothesis, with the other coefficients turned off.
The template fit approach improves the sensitivity.
}
\end{center}
\begin{center}
\includegraphics[width=\linewidth]{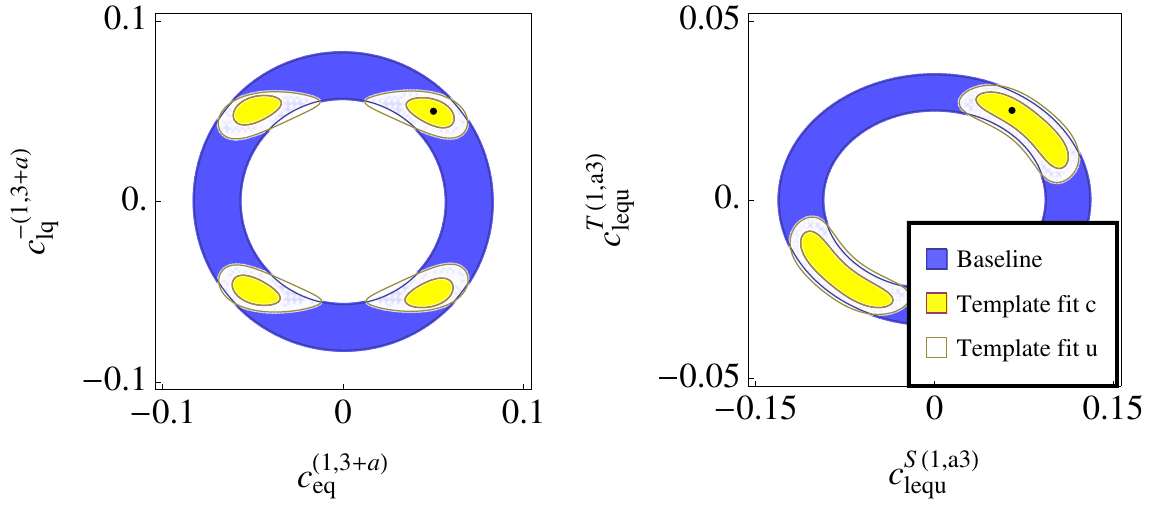}
\figcaption{\label{fig:discrimination}
Two-dimensional limits of the four-fermion coefficients, at 95\% CL,
with the other coefficients turned off.  Two hypotheses are considered.
Left: $c_{eq}^{(1,3+a)}=c_{lq}^{-(1,3+a)}=0.05$.
Right: $c_{lequ}^{S(1,a3)}=0.065$, $c_{lequ}^{T(1,a3)}=0.025$.
Both points are labeled by a black dot in the plots.
The template fit helps to pinpoint the coefficients. Better precision
is obtained for operators involving a charm-quark (i.e.~$a=2$).
}
\end{center}

\begin{center}
\includegraphics[width=0.8\linewidth]{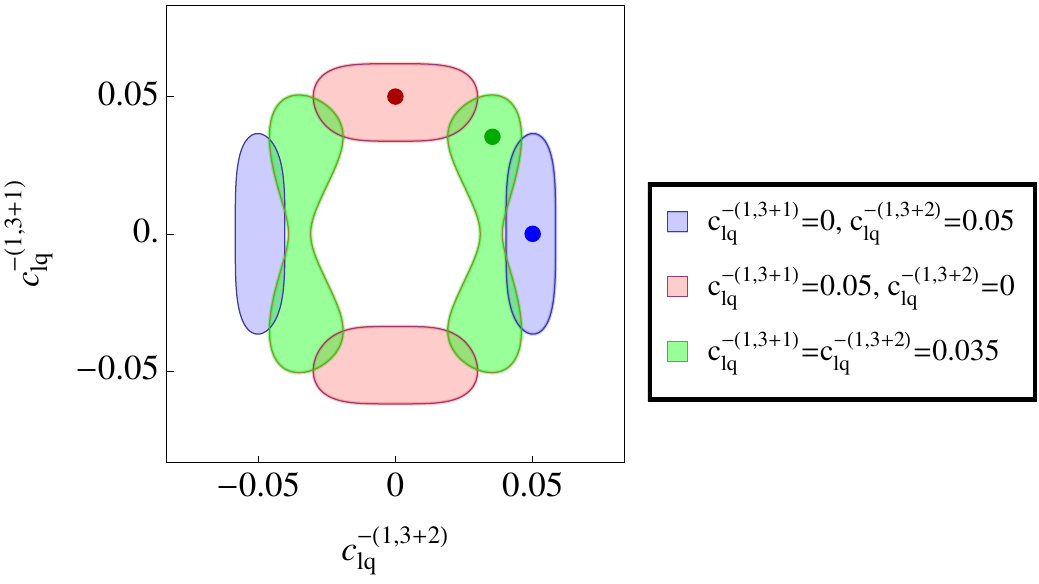}
\figcaption{\label{fig:uc}
Two-dimensional limits of the $c_{lq}^{-(1,3+a)}$ coefficients with $a=1$ and
$a=2$, at 95\% CL.  The other coefficients are turned off.  Three
hypotheses are considered.  The template fit helps to identify the light-quark
flavor involved in the FCN coupling.
}
\end{center}

A more comprehensive study can further improve these results in several
aspects.  The QCD correction of the four-fermion operators can be implemented
in the analysis, although we expect the correction to be similar to
the two-fermion case.  Kinematic features of the signals from different
operators can be fully exploited by using a multivariate analysis.
Alternatively, one could also construct the covariant matrix directly,
following the statistically optimal observable
\cite{Atwood:1991ka,Diehl:1993br}, which in theory guarantees the best sensitivity.
However, the nonlinear form of the cross-section in the parameter space and the
non-analytic nature of the detector effects need to be carefully dealt
with. The same approach has been used to study the FCN couplings at
the CLIC \cite{deBlas:2018mhx}, where the detector effects were taken into
account by an efficiency parameter. Finally, useful information may also come
from the study of flavor changing decay of the top quark, depending on the
possibility of an energy upgrade above the 350 GeV threshold, which in addition
could also provide access to the Higgs and gluon FCN couplings.  We defer these
studies to a future work.

\section{Conclusion}
\label{sec:conclusion}
The CEPC collider, proposed as a Higgs factory, is also an ideal machine to study
the flavor properties of the top quark.  The FCN interactions of the top quark
can be searched for in the single top production $e^+e^-\to tj$.  The results from
LEP2, Tevatron and LHC experiments suggest that a future lepton collider would
provide the best sensitivity for the four-fermion $eetq$ FCN interactions,
complementary to a hadron collider which mainly constrains the two-fermion FCN
interactions.  In this work, we derived the expected sensitivity at CEPC,
with an energy of 240 GeV and integrated luminosity of 5.6 ab$^{-1}$, of the full set of
56 FCN operators that are relevant for the $e^+e^-\to tj$ channel, and showed that
an improvement of about 1-2 orders of magnitude of the four-fermion
FCN couplings could be expected.  Our main results are displayed in
Figures~\ref{fig:individuallimits} and \ref{fig:2dlimits}, where one can
clearly see that a large fraction of the currently allowed FCN parameters could be tested
at CEPC.  We also showed that the capability of $c$-jet tagging at CEPC
further improves the sensitivity for the flavor-changing couplings between the top
and charm quarks.  In case a signature is established, we showed that kinematic
observables could be used to pinpoint the values of the coefficients, which in
turn would give information about the new physics behind the discovery.

\textbf{Note added:} After this work was posted on arXiv, Ref.~\cite{Liu:2019wmi}
appeared, where the authors discussed the expected limits of the four-fermion
coefficients at the Large Hadron-Electron Collider.  The results are
of the same order of magnitude
as what we gave in Figures~\ref{fig:individuallimits} and \ref{fig:individuallimits2}.

\bigskip
%\section*{Acknowledgements}
\textit{We would like to thank M.~Chala, B.~Fuks, G.~Durieux, Z.~Liang and H.-S.~Shao
for helpful discussions and suggestions.}

\end{multicols}%

\clearpage
\appendix
\counterwithin{figure}{section}

\section{Additional results}
\label{sec:ad}
\begin{multicols}{2}%

\begin{figure*}[ht]
\centering
\includegraphics[width=.8\linewidth]{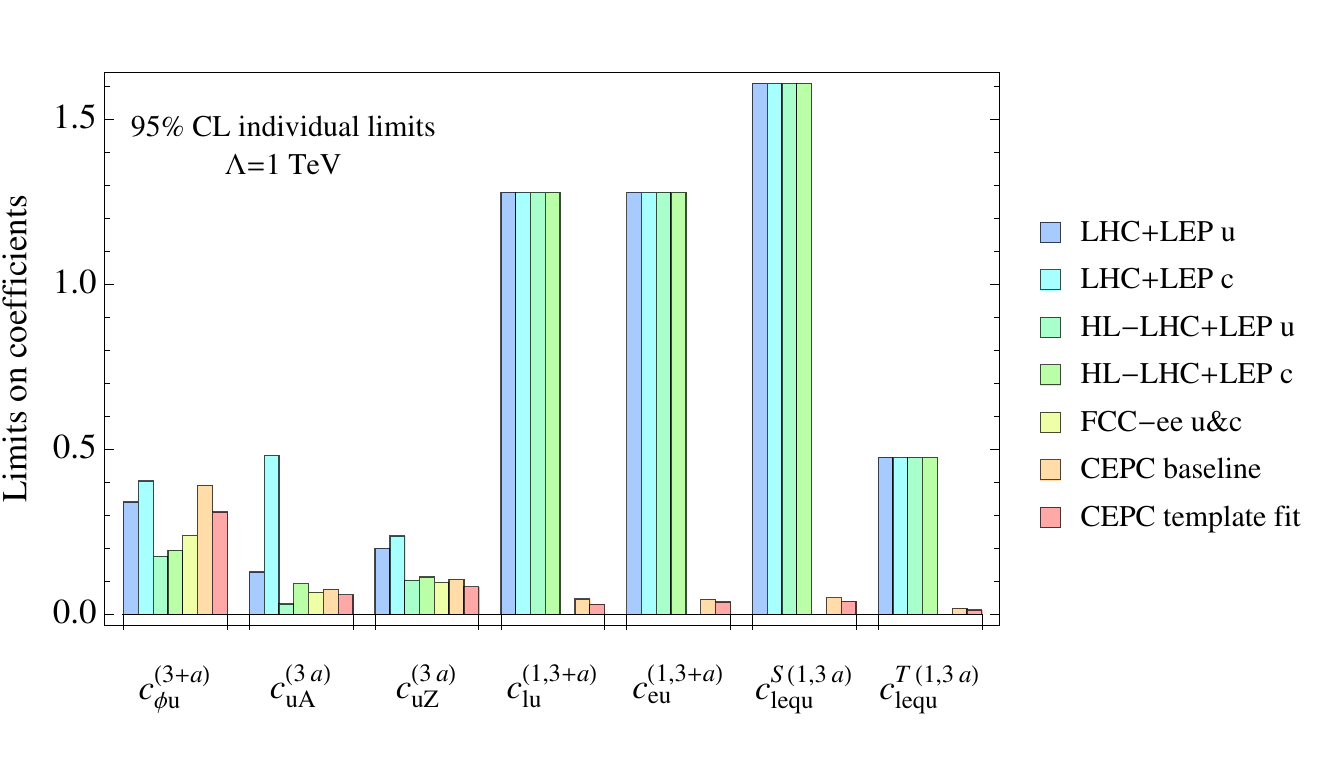}
\caption{\label{fig:individuallimits2}
The 95\% CL limits on individual coefficients in the second row of
Eq.~(\ref{eq:28}), expected from the CEPC, compared with the existing LHC+LEP2
bounds, and the projected limits from HL-LHC+LEP2 and from FCC-ee with 3 ab$^{-1}$
luminosity at 240 GeV (only for the first three coefficients), see
Refs.~\cite{Cerri:2018ypt,Khanpour:2014xla}.  Results for both generations
$a=1,2$ are displayed.  The orange column ``CEPC baseline'' is the expected
limits following our baseline analysis, which applies to both flavors (a=1,2).
The red column ``CEPC template fit'' uses the $c$-jet tagging in its signal
definition and only applies to $a=2$ operators.
}
\end{figure*}

Here, we list some additional results mentioned in the previous sections.
In Figures~\ref{fig:uzcomparison} and \ref{fig:eqcomparison}, we compare the
signals from $c_{uZ}^{(23)}$, $c_{uZ}^{(32)}$, $c_{uZ}^{I(23)}$, and from
$c_{eq}^{(1,3+2)}$, $c_{eu}^{(1,3+2)}$, $c_{eq}^{I(1,3+2)}$,
illustrating the relations between the coefficients in different rows of
Eq.~(\ref{eq:28}).  In Figure~\ref{fig:individuallimits2}, we show the
individual limits and prospects for the coefficients from the second row of
Eq.~(\ref{eq:28}), similar to Figure~\ref{fig:individuallimits}.  In
Figure~\ref{fig:2dlimits2}, we present the two-dimensional bound of the
two-fermion coefficient $c_{\varphi u}^{-(3+2)}$ and the four-fermion coefficient
$c_{eu}^{(1,3+2)}$, similar to Figure~\ref{fig:2dlimits}.  Finally, in
Figure~\ref{fig:discovery2}, we show the discovery limits of the coefficients of
the second row of Eq.~(\ref{eq:28}), similar to Figure~\ref{fig:discovery2}.

\end{multicols}%
\begin{multicols}{2}%

\begin{center}
\includegraphics[width=.9\linewidth]{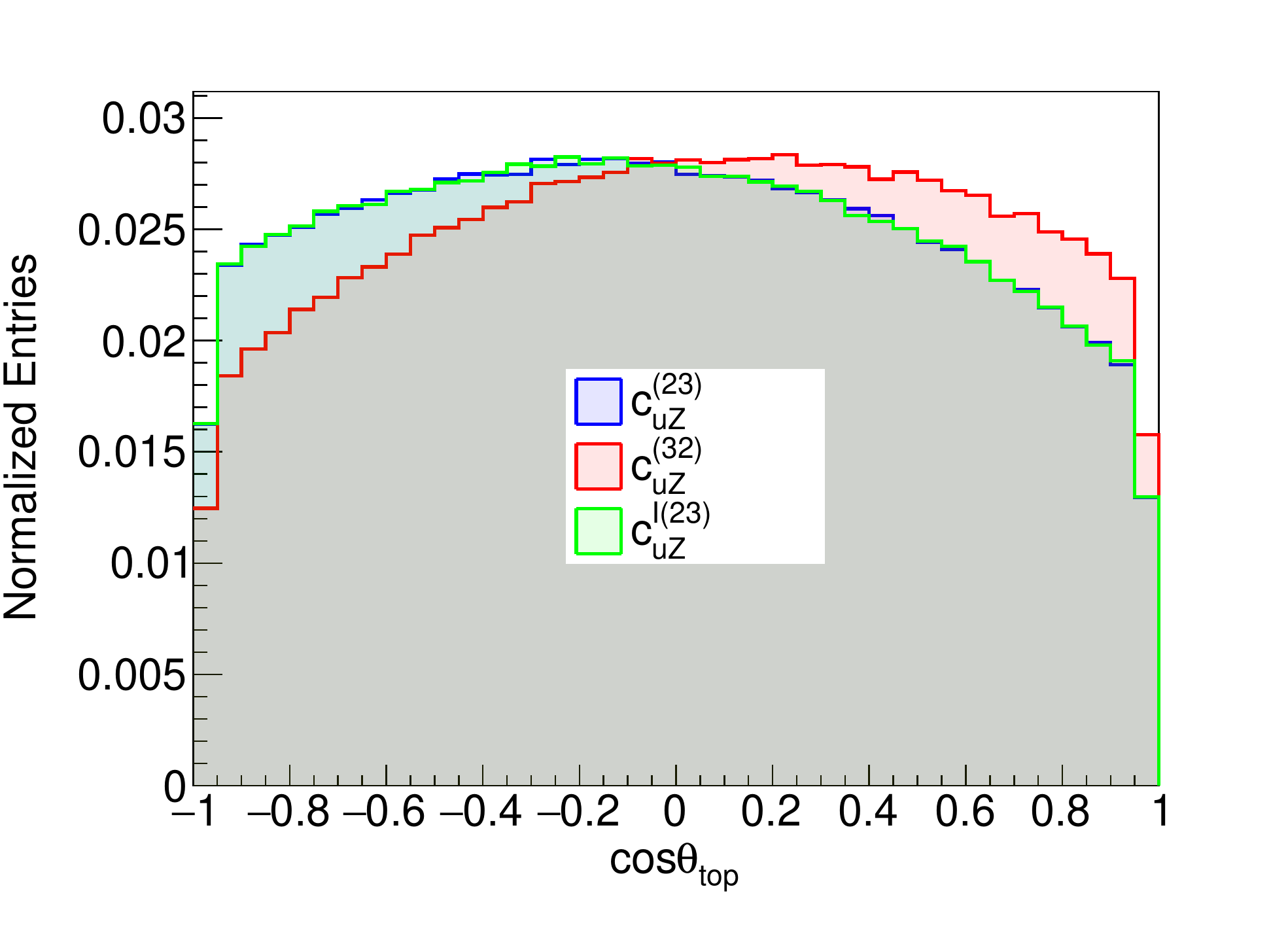}
\includegraphics[width=.9\linewidth]{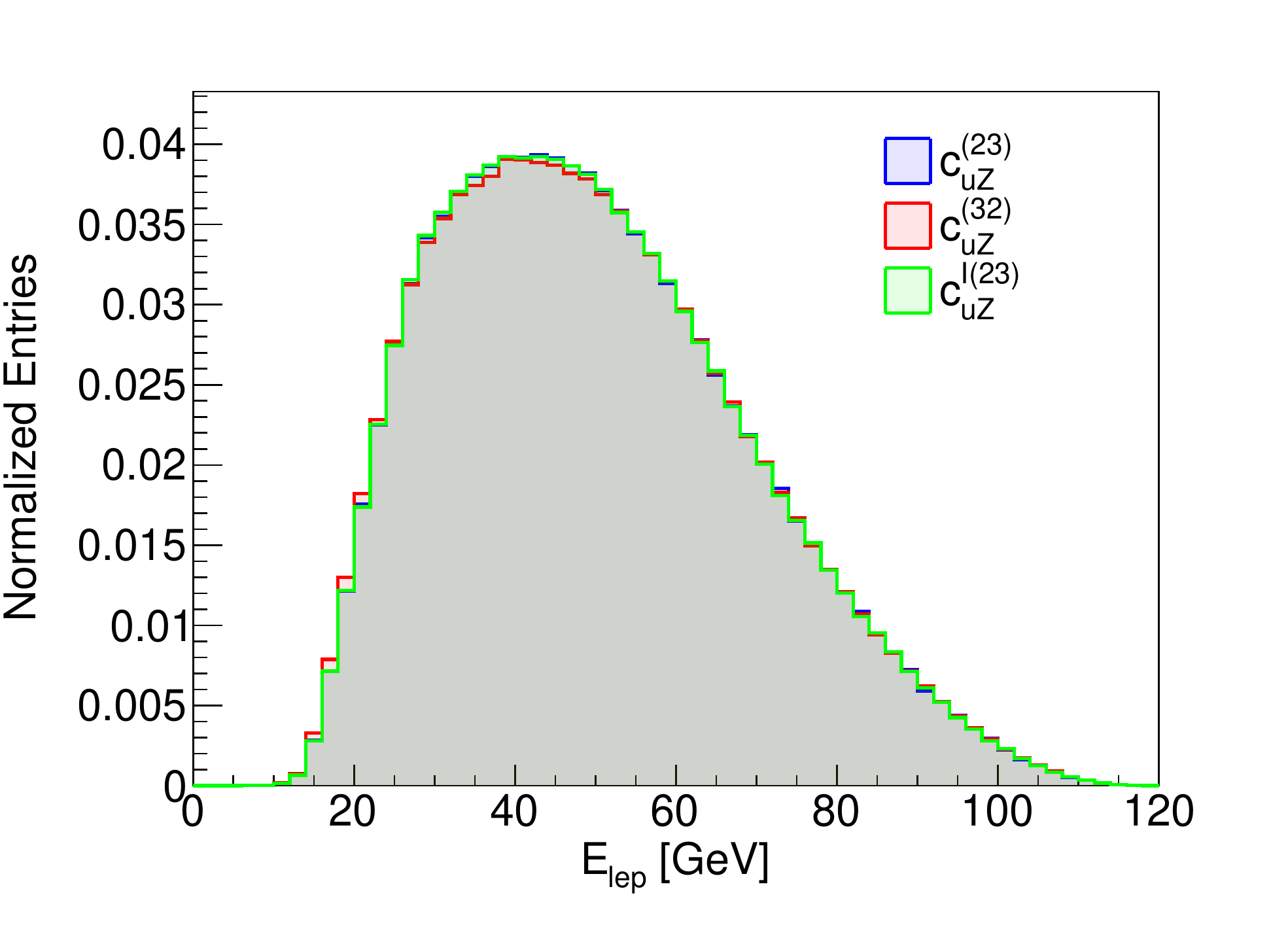}
\includegraphics[width=.9\linewidth]{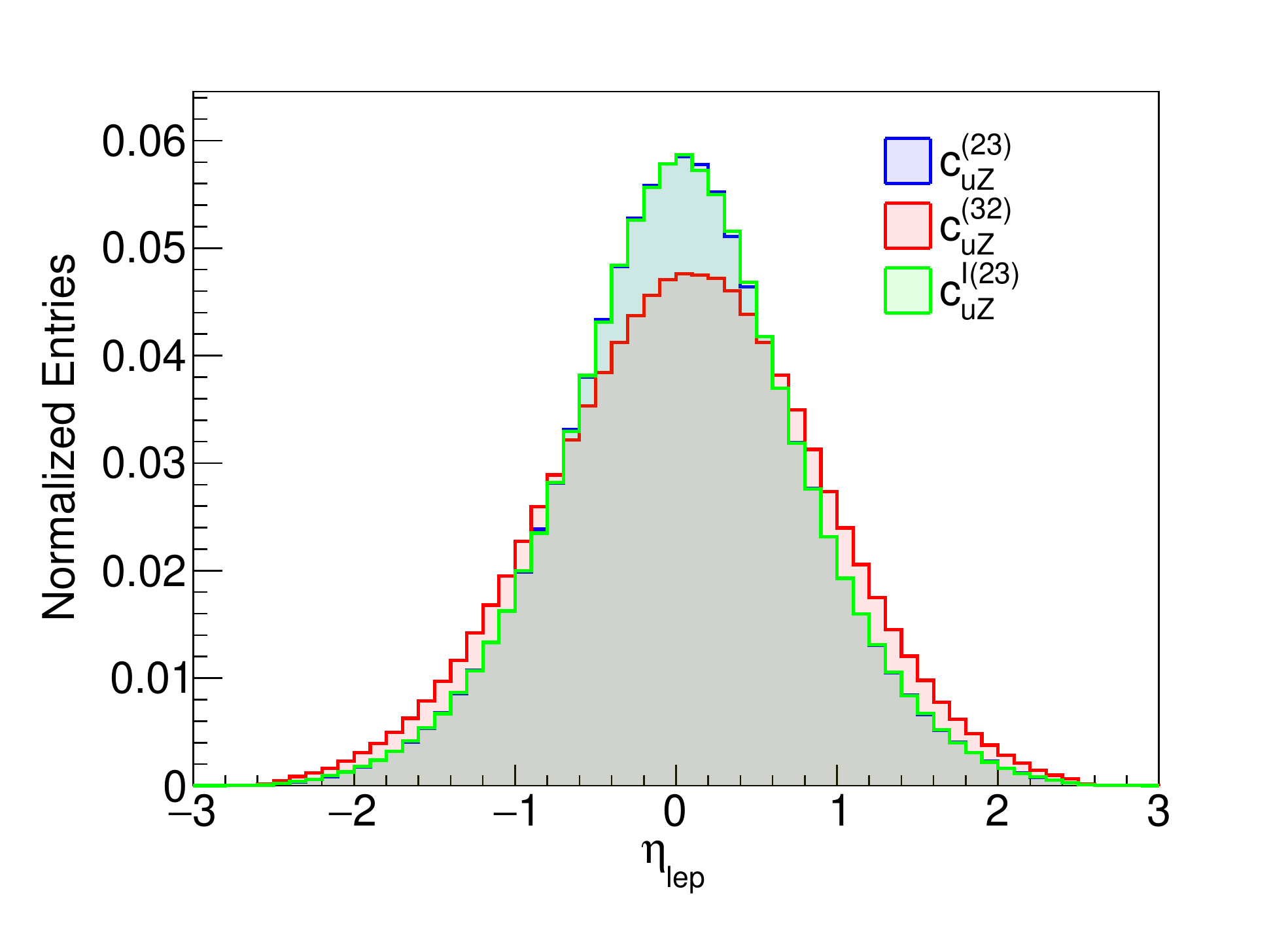}
\figcaption{\label{fig:uzcomparison}
	Signals from $c_{uZ}^{(23)}$, $c_{uZ}^{(32)}$ and
$c_{uZ}^{I(23)}$ at the parton level. Distributions of the scattering angle,
the lepton energy, and the lepton pseudorapidity are compared.
}
\end{center}

\begin{center}
\includegraphics[width=.9\linewidth]{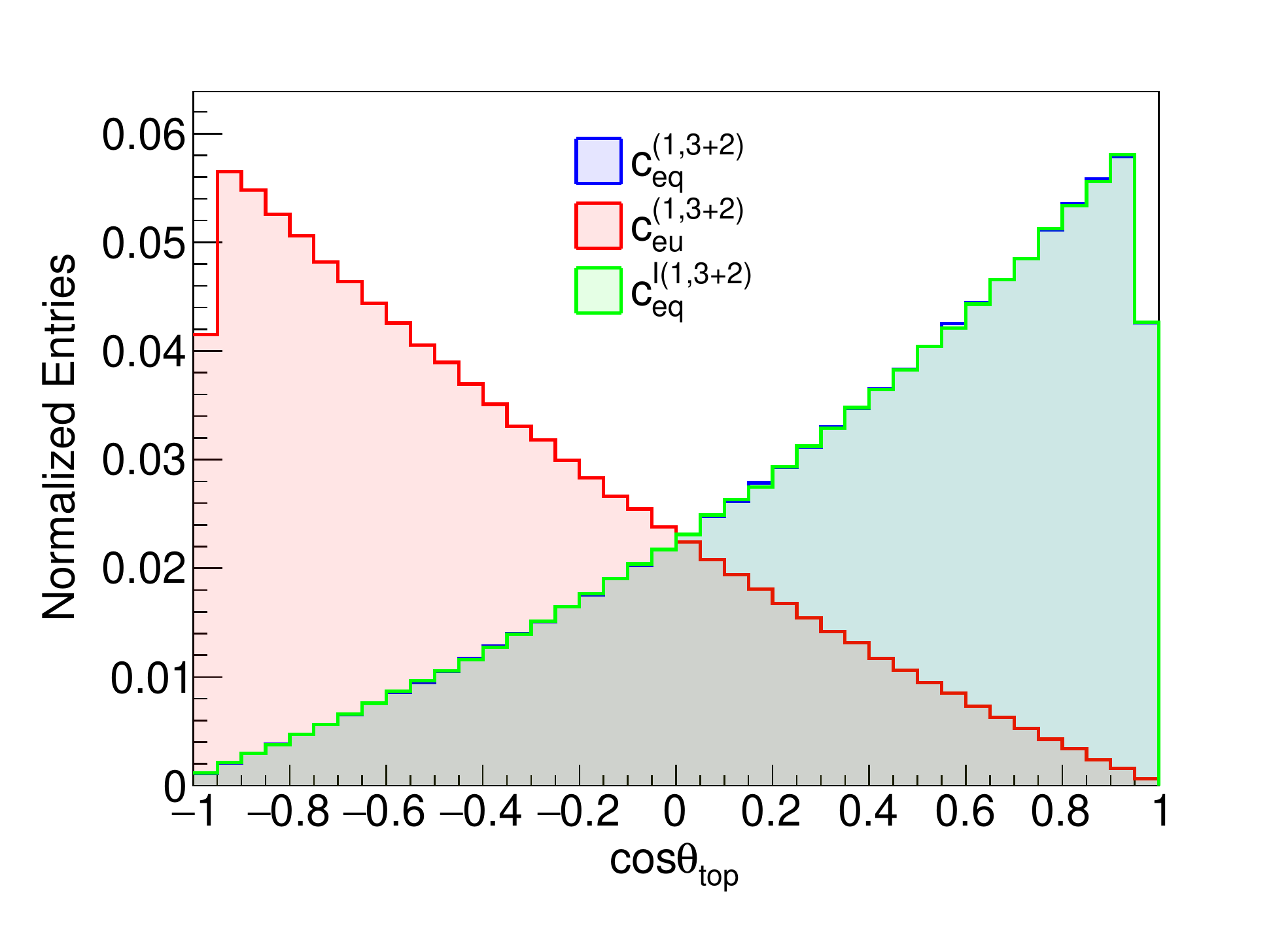}
\includegraphics[width=.9\linewidth]{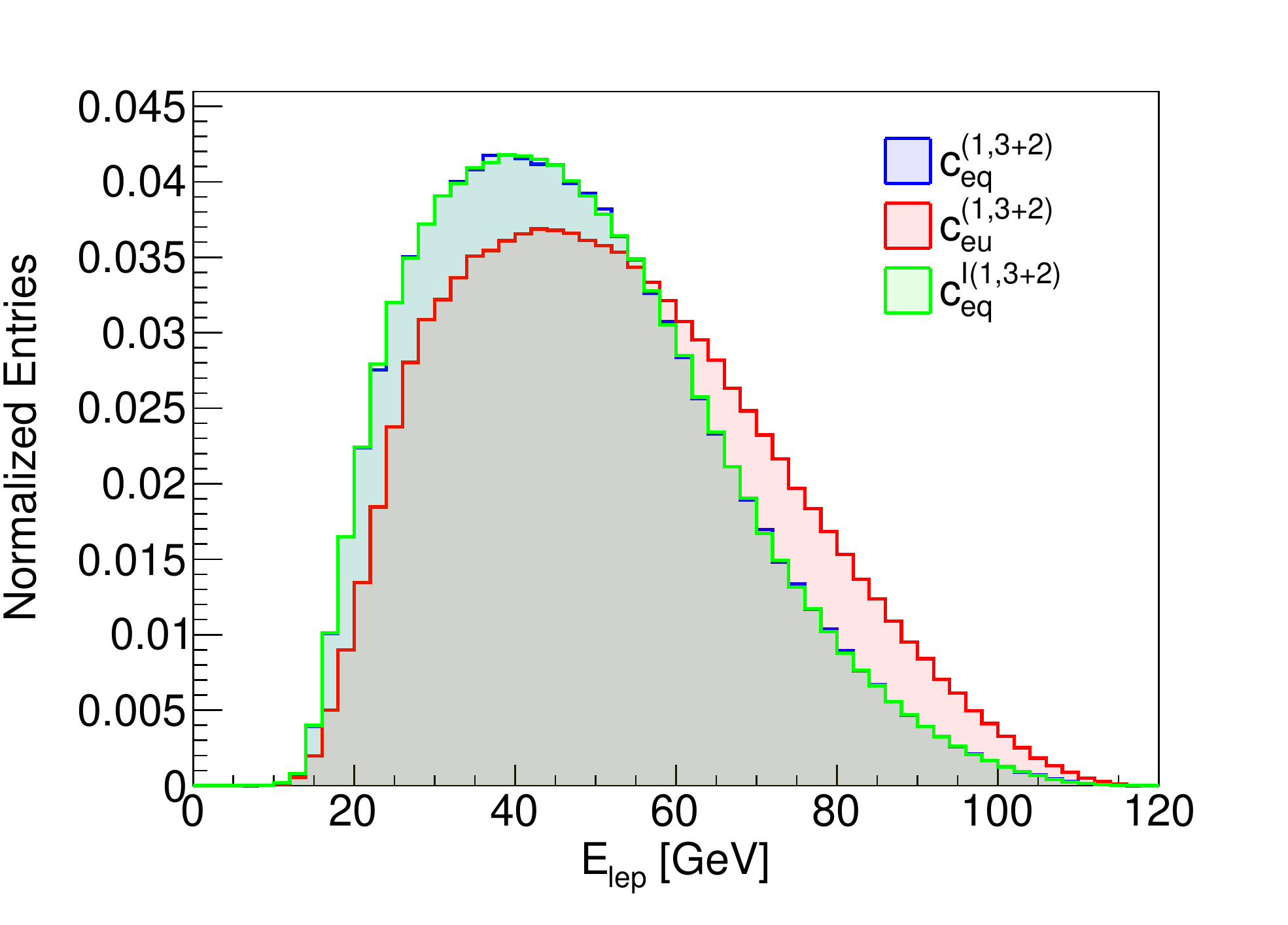}
\includegraphics[width=.9\linewidth]{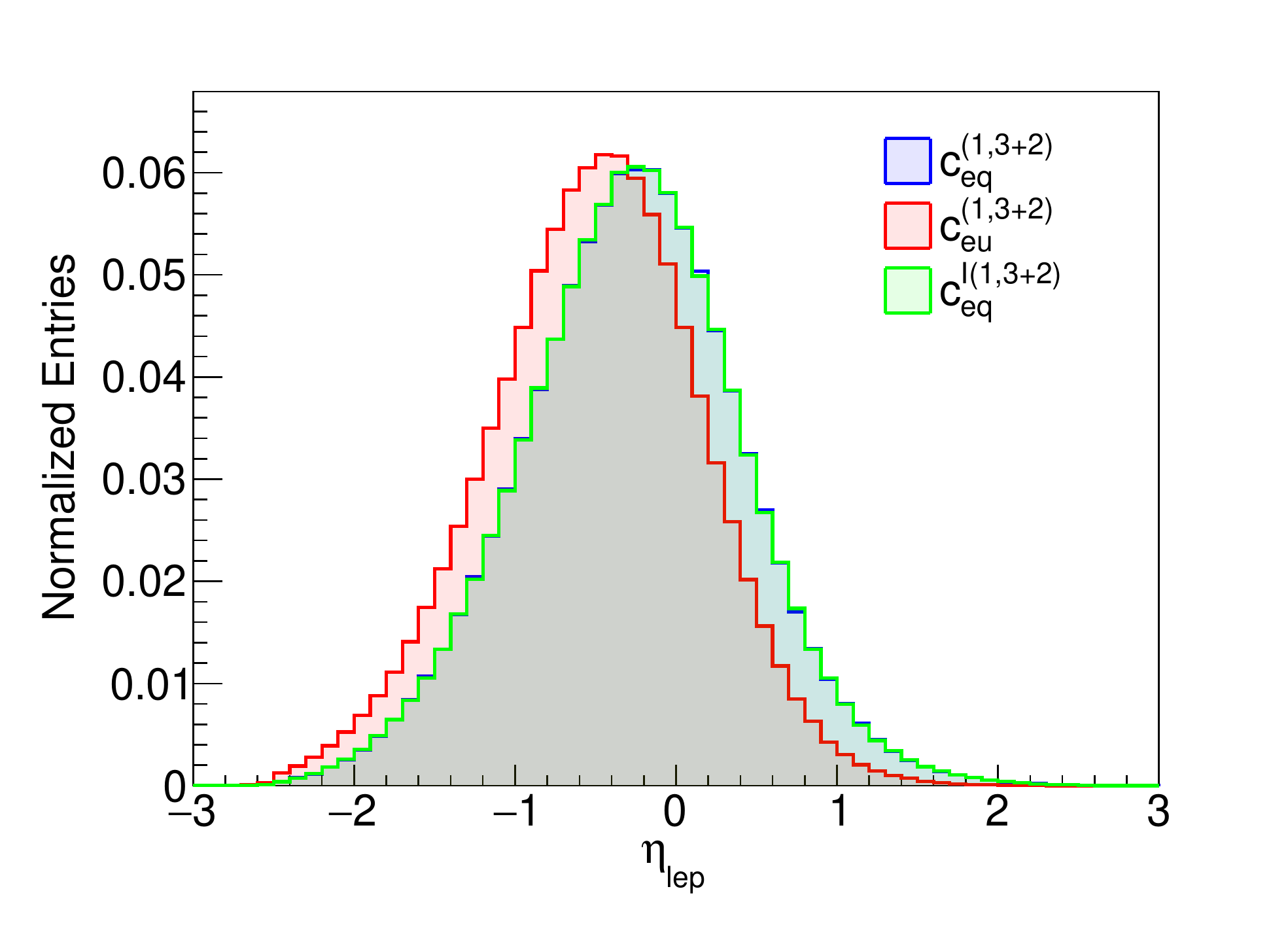}
\figcaption{\label{fig:eqcomparison}
	Signals from $c_{eq}^{(1,3+2)}$, $c_{eu}^{(1,3+2)}$ and
$c_{eq}^{I(1,3+2)}$ at the parton level. Distributions of the scattering angle,
the lepton energy, and the lepton pseudorapidity are compared.
}
\end{center}

\begin{center}
\includegraphics[width=.75\linewidth]{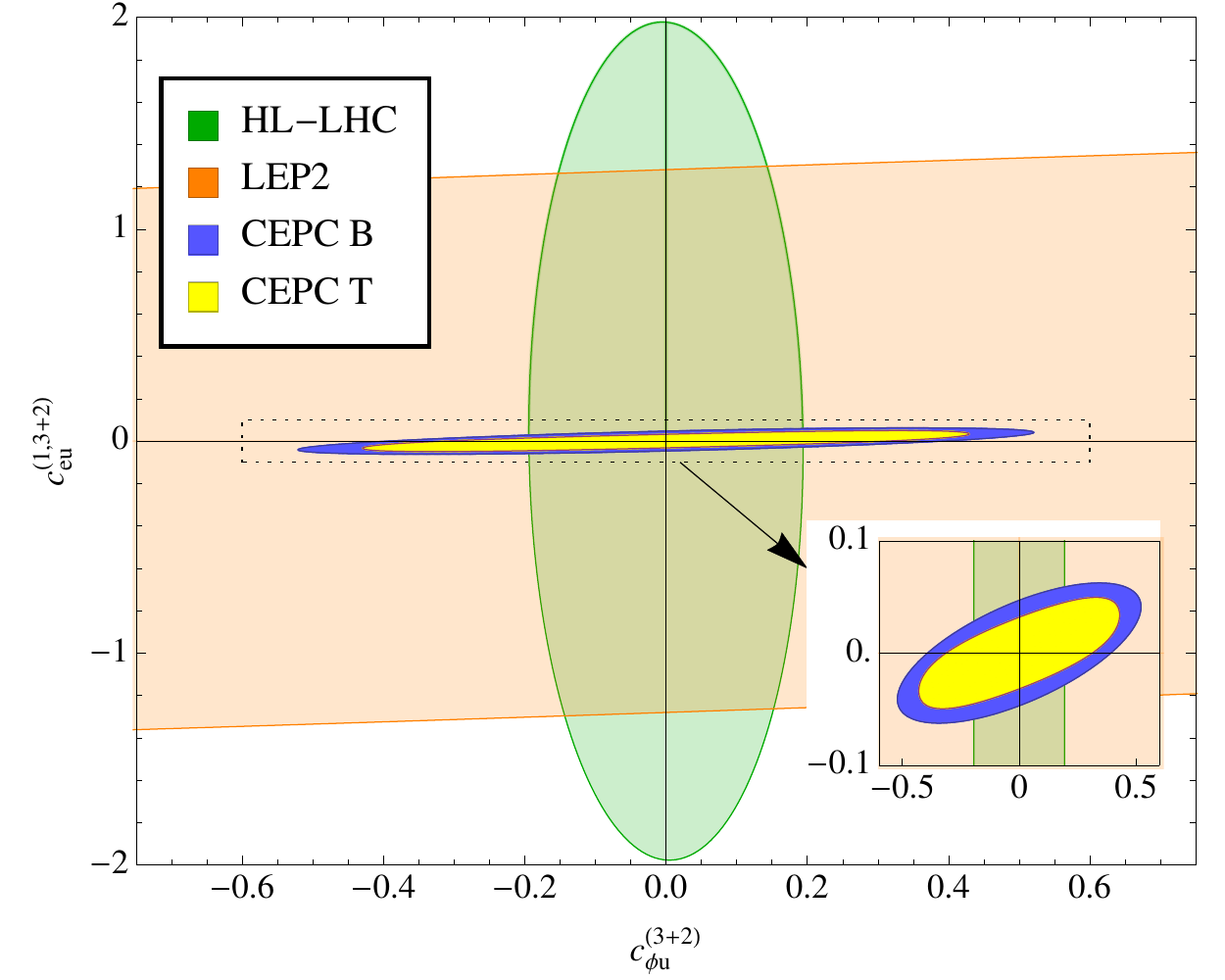}
\figcaption{\label{fig:2dlimits2}
Two-dimensional bounds on a two-fermion coefficient
$c_{\varphi u}^{-(3+2)}$ and a four-fermion coefficient $c_{eu}^{(1,3+2)}$,
at the 95\% CL.
Other operators are fixed at 0. The allowed regions from HL-LHC and LEP2
are similar to Figure~59 in Section 8.1 of Ref.~\cite{Cerri:2018ypt}, except for that
there all coefficients are marginalized over. The blue region (``CEPC B'') is
the bound expected from the CEPC, following our baseline analysis.  The
yellow region (``CEPC T'') is obtained with a template fit approach,
see more discussions in Section~\ref{sec:results}.
}
\end{center}

\begin{center}
\includegraphics[width=.95\linewidth]{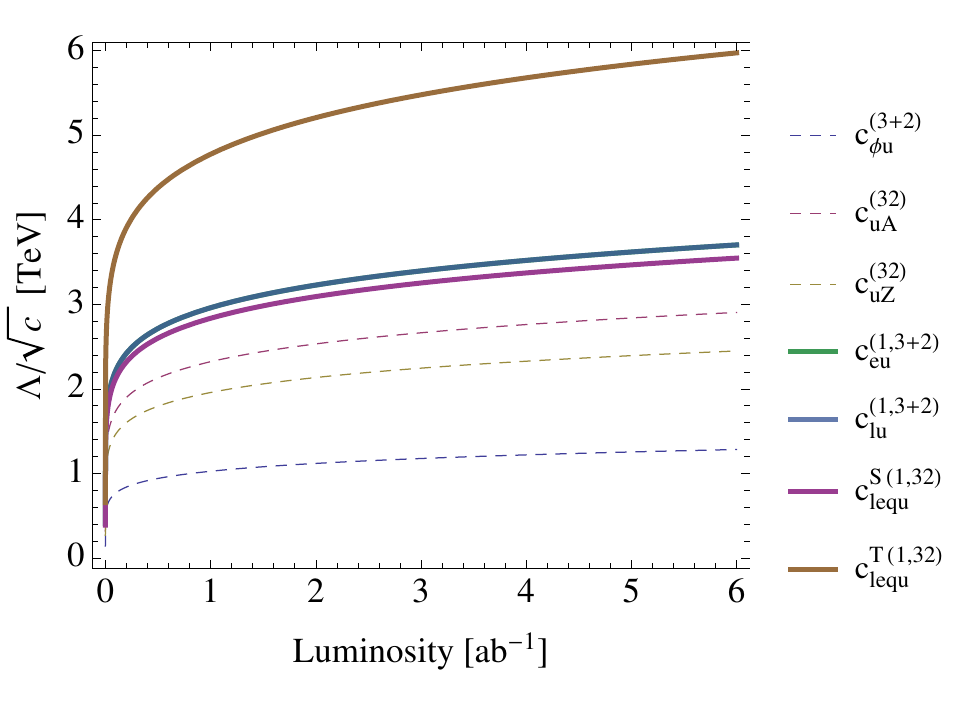}
\figcaption{\label{fig:discovery2}
Five-sigma discovery limit of $\Lambda/\sqrt{c}$, which is roughly the scale
of new physics, for coefficients in the second row of Eq.~(\ref{eq:28}), as
function of integrated luminosity at the CEPC.
}
\end{center}

%%%%%%%%%%%%%%%%%%%%%%%%%%%%%%%%%%%%%%%
\vspace{3mm}\nopagebreak[4]

\end{multicols}%
\begin{multicols}{2}%

\end{multicols}

\begin{thebibliography}{90}

%\cite{Aad:2012tfa}
\bibitem{Aad:2012tfa} 
  G.~Aad {\it et al.} [ATLAS Collaboration],
  %``Observation of a new particle in the search for the Standard Model Higgs boson with the ATLAS detector at the LHC,''
  Phys.\ Lett.\ B {\bf 716}, 1 (2012)
  doi:10.1016/j.physletb.2012.08.020
  [arXiv:1207.7214 [hep-ex]].
  %%CITATION = doi:10.1016/j.physletb.2012.08.020;%%
  %8671 citations counted in INSPIRE as of 22 Jul 2018


%\cite{Chatrchyan:2012xdj}
\bibitem{Chatrchyan:2012xdj} 
  S.~Chatrchyan {\it et al.} [CMS Collaboration],
  %``Observation of a new boson at a mass of 125 GeV with the CMS experiment at the LHC,''
  Phys.\ Lett.\ B {\bf 716}, 30 (2012)
  doi:10.1016/j.physletb.2012.08.021
  [arXiv:1207.7235 [hep-ex]].
  %%CITATION = doi:10.1016/j.physletb.2012.08.021;%%
  %8474 citations counted in INSPIRE as of 22 Jul 2018

%\cite{Cepeda:2019klc}
\bibitem{Cepeda:2019klc} 
  M.~Cepeda {\it et al.} [HL/HE WG2 group],
  %``Higgs Physics at the HL-LHC and HE-LHC,''
  arXiv:1902.00134 [hep-ph].
  %%CITATION = ARXIV:1902.00134;%%
  %52 citations counted in INSPIRE as of 23 Jul 2019

%\cite{CEPC-SPPCStudyGroup:2015csa}
\bibitem{CEPC-SPPCStudyGroup:2015csa} 
  M.~Ahmad {\it et al.},
  %``CEPC-SPPC Preliminary Conceptual Design Report. 1. Physics and Detector,''
  IHEP-CEPC-DR-2015-01, IHEP-TH-2015-01, IHEP-EP-2015-01.
  %%CITATION = IHEP-CEPC-DR-2015-01, IHEP-TH-2015-01, IHEP-EP-2015-01;%%
  %162 citations counted in INSPIRE as of 02 May 2019

\bibitem{CEPCStudyGroup:2018ghi} 
  J.~Guimaraes da Costa {\it et al.} [CEPC Study Group],
  %``CEPC Conceptual Design Report: Volume 2 - Physics & Detector,''
  arXiv:1811.10545 [hep-ex].
  %%CITATION = ARXIV:1811.10545;%%
  %39 citations counted in INSPIRE as of 02 May 2019

  %\cite{Hill:2002ap}
\bibitem{Hill:2002ap} 
  C.~T.~Hill and E.~H.~Simmons,
  %``Strong dynamics and electroweak symmetry breaking,''
  Phys.\ Rept.\  {\bf 381}, 235 (2003)
  Erratum: [Phys.\ Rept.\  {\bf 390}, 553 (2004)]
  doi:10.1016/S0370-1573(03)00140-6
  [hep-ph/0203079].
  %%CITATION = doi:10.1016/S0370-1573(03)00140-6;%%
  %993 citations counted in INSPIRE as of 31 Jul 2019

%\cite{Vryonidou:2018eyv}
\bibitem{Vryonidou:2018eyv} 
  E.~Vryonidou and C.~Zhang,
  %``Dimension-six electroweak top-loop effects in Higgs production and decay,''
  JHEP {\bf 1808}, 036 (2018)
  doi:10.1007/JHEP08(2018)036
  [arXiv:1804.09766 [hep-ph]].
  %%CITATION = doi:10.1007/JHEP08(2018)036;%%
  %17 citations counted in INSPIRE as of 02 May 2019

%\cite{Boselli:2018zxr}
\bibitem{Boselli:2018zxr} 
  S.~Boselli, R.~Hunter and A.~Mitov,
  %``Prospects for the determination of the top-quark Yukawa coupling at future $e^+e^-$ colliders,''
  arXiv:1805.12027 [hep-ph].
  %%CITATION = ARXIV:1805.12027;%%
  %7 citations counted in INSPIRE as of 23 May 2019

%\cite{Durieux:2018ggn}
\bibitem{Durieux:2018ggn} 
  G.~Durieux, J.~Gu, E.~Vryonidou and C.~Zhang,
  %``Probing top-quark couplings indirectly at Higgs factories,''
  Chin.\ Phys.\ C {\bf 42}, no. 12, 123107 (2018)
  doi:10.1088/1674-1137/42/12/123107
  [arXiv:1809.03520 [hep-ph]].
  %%CITATION = doi:10.1088/1674-1137/42/12/123107;%%
  %5 citations counted in INSPIRE as of 02 May 2019

%\cite{Glashow:1970gm}
\bibitem{Glashow:1970gm} 
  S.~L.~Glashow, J.~Iliopoulos and L.~Maiani,
  %``Weak Interactions with Lepton-Hadron Symmetry,''
  Phys.\ Rev.\ D {\bf 2}, 1285 (1970).
  doi:10.1103/PhysRevD.2.1285
  %%CITATION = doi:10.1103/PhysRevD.2.1285;%%
  %5743 citations counted in INSPIRE as of 02 May 2019
  

  

%\cite{Abe:1997fz}
\bibitem{Abe:1997fz} 
  F.~Abe {\it et al.} [CDF Collaboration],
  %``Search for flavor-changing neutral current decays of the top quark in $p \bar{p}$ collisions at $\sqrt{s} = 1.8$ TeV,''
  Phys.\ Rev.\ Lett.\  {\bf 80}, 2525 (1998).
  doi:10.1103/PhysRevLett.80.2525
  %%CITATION = doi:10.1103/PhysRevLett.80.2525;%%
  %228 citations counted in INSPIRE as of 02 May 2019



%\cite{Aaltonen:2008ac}
\bibitem{Aaltonen:2008ac} 
  T.~Aaltonen {\it et al.} [CDF Collaboration],
  %``Search for the Flavor Changing Neutral Current Decay $t \to Zq$ in $p\bar{p}$ Collisions at $\sqrt{s} = 1.96$ TeV,''
  Phys.\ Rev.\ Lett.\  {\bf 101}, 192002 (2008)
  doi:10.1103/PhysRevLett.101.192002
  [arXiv:0805.2109 [hep-ex]].
  %%CITATION = doi:10.1103/PhysRevLett.101.192002;%%
  %102 citations counted in INSPIRE as of 02 May 2019



%\cite{Aaltonen:2009ef}
\bibitem{Aaltonen:2009ef} 
  T.~Aaltonen {\it et al.} [CDF Collaboration],
  %``Search for the Neutral Current Top Quark Decay t ---> Zc Using Ratio of Z-Boson + 4 Jets to W-Boson + 4 Jets Production,''
  Phys.\ Rev.\ D {\bf 80}, 052001 (2009)
  doi:10.1103/PhysRevD.80.052001
  [arXiv:0905.0277 [hep-ex]].
  %%CITATION = doi:10.1103/PhysRevD.80.052001;%%
  %5 citations counted in INSPIRE as of 02 May 2019



%\cite{Abazov:2011qf}
\bibitem{Abazov:2011qf} 
  V.~M.~Abazov {\it et al.} [D0 Collaboration],
  %``Search for flavor changing neutral currents in decays of top quarks,''
  Phys.\ Lett.\ B {\bf 701}, 313 (2011)
  doi:10.1016/j.physletb.2011.06.014
  [arXiv:1103.4574 [hep-ex]].
  %%CITATION = doi:10.1016/j.physletb.2011.06.014;%%
  %69 citations counted in INSPIRE as of 02 May 2019

%decay LHC


  %\cite{Aaboud:2018nyl}
\bibitem{Aaboud:2018nyl} 
  M.~Aaboud {\it et al.} [ATLAS Collaboration],
  %``Search for flavour-changing neutral current top-quark decays $t\to qZ$ in proton-proton collisions at $\sqrt{s}=13$ TeV with the ATLAS detector,''
  JHEP {\bf 1807}, 176 (2018)
  doi:10.1007/JHEP07(2018)176
  [arXiv:1803.09923 [hep-ex]].
  %%CITATION = doi:10.1007/JHEP07(2018)176;%%
  %16 citations counted in INSPIRE as of 04 May 2019


%\cite{Aad:2015uza}
\bibitem{Aad:2015uza} 
  G.~Aad {\it et al.} [ATLAS Collaboration],
  %``Search for flavour-changing neutral current top-quark decays to $qZ$ in $pp$ collision data collected with the ATLAS detector at $\sqrt s =8$  TeV,''
  Eur.\ Phys.\ J.\ C {\bf 76}, no. 1, 12 (2016)
  doi:10.1140/epjc/s10052-015-3851-5
  [arXiv:1508.05796 [hep-ex]].
  %%CITATION = doi:10.1140/epjc/s10052-015-3851-5;%%
  %48 citations counted in INSPIRE as of 04 May 2019


%\cite{Aad:2012ij}
\bibitem{Aad:2012ij} 
  G.~Aad {\it et al.} [ATLAS Collaboration],
  %``A search for flavour changing neutral currents in top-quark decays in $pp$ collision data collected with the ATLAS detector at $\sqrt{s}=7$ TeV,''
  JHEP {\bf 1209}, 139 (2012)
  doi:10.1007/JHEP09(2012)139
  [arXiv:1206.0257 [hep-ex]].
  %%CITATION = doi:10.1007/JHEP09(2012)139;%%
  %75 citations counted in INSPIRE as of 02 May 2019



%\cite{ATLAS:2011mka}
\bibitem{ATLAS:2011mka} 
  [ATLAS Collaboration],
  %``A search for Flavour Changing Neutral Currents in Top Quark Decays t-> qZ at sqrt{s}=7 TeV in 0.70 fb$^{-1}$ of pp collision data collected with the ATLAS Detector,''
  ATLAS-CONF-2011-154.
  %%CITATION = ATLAS-CONF-2011-154;%%
  %12 citations counted in INSPIRE as of 02 May 2019



%\cite{ATLAS:2011jga}
\bibitem{ATLAS:2011jga} 
  [ATLAS Collaboration],
  %``Search for FCNC top quark processes at 7 TeV with the ATLAS detector,''
  ATLAS-CONF-2011-061.
  %%CITATION = ATLAS-CONF-2011-061;%%
  %20 citations counted in INSPIRE as of 02 May 2019


  %\cite{CMS:2017twu}
\bibitem{CMS:2017twu} 
  CMS Collaboration [CMS Collaboration],
  %``Search for flavour changing neutral currents in top quark production and decays with three-lepton final state using the data collected at sqrt(s) = 13 TeV,''
  CMS-PAS-TOP-17-017.
  %%CITATION = CMS-PAS-TOP-17-017;%%
  %9 citations counted in INSPIRE as of 04 May 2019


%\cite{Chatrchyan:2013nwa}
\bibitem{Chatrchyan:2013nwa} 
  S.~Chatrchyan {\it et al.} [CMS Collaboration],
  %``Search for Flavor-Changing Neutral Currents in Top-Quark Decays $t \to Zq$ in $pp$ Collisions at $\sqrt{s}=8$  TeV,''
  Phys.\ Rev.\ Lett.\  {\bf 112}, no. 17, 171802 (2014)
  doi:10.1103/PhysRevLett.112.171802
  [arXiv:1312.4194 [hep-ex]].
  %%CITATION = doi:10.1103/PhysRevLett.112.171802;%%
  %96 citations counted in INSPIRE as of 02 May 2019



%\cite{Chatrchyan:2012hqa}
\bibitem{Chatrchyan:2012hqa} 
  S.~Chatrchyan {\it et al.} [CMS Collaboration],
  %``Search for flavor changing neutral currents in top quark decays in pp collisions at 7 TeV,''
  Phys.\ Lett.\ B {\bf 718}, 1252 (2013)
  doi:10.1016/j.physletb.2012.12.045
  [arXiv:1208.0957 [hep-ex]].
  %%CITATION = doi:10.1016/j.physletb.2012.12.045;%%
  %54 citations counted in INSPIRE as of 02 May 2019

%t->qh at LHC

%\cite{Aaboud:2018oqm}
\bibitem{Aaboud:2018oqm} 
  M.~Aaboud {\it et al.} [ATLAS Collaboration],
  %``Search for top-quark decays $t \rightarrow Hq$ with 36 fb$^{-1}$ of $pp$ collision data at $\sqrt{s}=13$ TeV with the ATLAS detector,''
  %Submitted to: JHEP
  [arXiv:1812.11568 [hep-ex]].
  %%CITATION = ARXIV:1812.11568;%%
  %5 citations counted in INSPIRE as of 04 May 2019


  %\cite{Aaboud:2018pob}
\bibitem{Aaboud:2018pob} 
  M.~Aaboud {\it et al.} [ATLAS Collaboration],
  %``Search for flavor-changing neutral currents in top quark decays $t\to Hc$ and $t \to Hu$ in multilepton final states in proton-proton collisions at $\sqrt{s}= 13$ TeV with the ATLAS detector,''
  Phys.\ Rev.\ D {\bf 98}, no. 3, 032002 (2018)
  doi:10.1103/PhysRevD.98.032002
  [arXiv:1805.03483 [hep-ex]].
  %%CITATION = doi:10.1103/PhysRevD.98.032002;%%
  %15 citations counted in INSPIRE as of 04 May 2019



%\cite{Sirunyan:2017uae}
\bibitem{Sirunyan:2017uae} 
  A.~M.~Sirunyan {\it et al.} [CMS Collaboration],
  %``Search for the flavor-changing neutral current interactions of the top quark and the Higgs boson which decays into a pair of b quarks at $\sqrt{s}=$ 13 TeV,''
  JHEP {\bf 1806}, 102 (2018)
  doi:10.1007/JHEP06(2018)102
  [arXiv:1712.02399 [hep-ex]].
  %%CITATION = doi:10.1007/JHEP06(2018)102;%%
  %22 citations counted in INSPIRE as of 04 May 2019


  %\cite{Aaboud:2017mfd}
\bibitem{Aaboud:2017mfd} 
  M.~Aaboud {\it et al.} [ATLAS Collaboration],
  %``Search for top quark decays $t\rightarrow qH$, with $H\to\gamma\gamma$, in $\sqrt{s}=13$ TeV $pp$ collisions using the ATLAS detector,''
  JHEP {\bf 1710}, 129 (2017)
  doi:10.1007/JHEP10(2017)129
  [arXiv:1707.01404 [hep-ex]].
  %%CITATION = doi:10.1007/JHEP10(2017)129;%%
  %43 citations counted in INSPIRE as of 04 May 2019


  %\cite{Khachatryan:2016atv}
\bibitem{Khachatryan:2016atv} 
  V.~Khachatryan {\it et al.} [CMS Collaboration],
  %``Search for top quark decays via Higgs-boson-mediated flavor-changing neutral currents in pp collisions at $ \sqrt{s}=8 $ TeV,''
  JHEP {\bf 1702}, 079 (2017)
  doi:10.1007/JHEP02(2017)079
  [arXiv:1610.04857 [hep-ex]].
  %%CITATION = doi:10.1007/JHEP02(2017)079;%%
  %42 citations counted in INSPIRE as of 04 May 2019
 

  %\cite{Aad:2015pja}
\bibitem{Aad:2015pja} 
  G.~Aad {\it et al.} [ATLAS Collaboration],
  %``Search for flavour-changing neutral current top quark decays $t\to Hq$ in $pp$ collisions at $\sqrt{s}=8$ TeV with the ATLAS detector,''
  JHEP {\bf 1512}, 061 (2015)
  doi:10.1007/JHEP12(2015)061
  [arXiv:1509.06047 [hep-ex]].
  %%CITATION = doi:10.1007/JHEP12(2015)061;%%
  %75 citations counted in INSPIRE as of 04 May 2019


%\cite{CMS:2015qhe}
\bibitem{CMS:2015qhe} 
  CMS Collaboration [CMS Collaboration],
  %``Search for the Flavor-Changing Neutral Current Decay ${\rm t} \to {\rm qH}$ Where the Higgs Decays to $\rm b \overline b$ Pairs at $\sqrt{s} = 8$ TeV,''
  CMS-PAS-TOP-14-020.
  %%CITATION = CMS-PAS-TOP-14-020;%%
  %16 citations counted in INSPIRE as of 04 May 2019


%\cite{Aad:2014dya}
\bibitem{Aad:2014dya} 
  G.~Aad {\it et al.} [ATLAS Collaboration],
  %``Search for top quark decays $t \to qH$ with $H \to \gamma\gamma$ using the ATLAS detector,''
  JHEP {\bf 1406}, 008 (2014)
  doi:10.1007/JHEP06(2014)008
  [arXiv:1403.6293 [hep-ex]].
  %%CITATION = doi:10.1007/JHEP06(2014)008;%%
  %102 citations counted in INSPIRE as of 02 May 2019



%\cite{CMS:2014qxa}
\bibitem{CMS:2014qxa} 
  CMS Collaboration [CMS Collaboration],
  %``Combined multilepton and diphoton limit on t to cH,''
  CMS-PAS-HIG-13-034.
  %%CITATION = CMS-PAS-HIG-13-034;%%
  %62 citations counted in INSPIRE as of 02 May 2019


%\cite{TheATLAScollaboration:2013nia}
\bibitem{TheATLAScollaboration:2013nia} 
  The ATLAS collaboration [ATLAS Collaboration],
  %``Search for flavor-changing neutral currents in $t\rightarrow cH$, with $H\to\gamma\gamma$, and limit on the tcH coupling,''
  ATLAS-CONF-2013-081.
  %%CITATION = ATLAS-CONF-2013-081;%%
  %27 citations counted in INSPIRE as of 02 May 2019


%\cite{Aaltonen:2008qr}
\bibitem{Aaltonen:2008qr} 
  T.~Aaltonen {\it et al.} [CDF Collaboration],
  %``Search for top-quark production via flavor-changing neutral currents in W+1 jet events at CDF,''
  Phys.\ Rev.\ Lett.\  {\bf 102}, 151801 (2009)
  doi:10.1103/PhysRevLett.102.151801
  [arXiv:0812.3400 [hep-ex]].
  %%CITATION = doi:10.1103/PhysRevLett.102.151801;%%
  %77 citations counted in INSPIRE as of 02 May 2019


  %\cite{Aad:2015gea}
\bibitem{Aad:2015gea} 
  G.~Aad {\it et al.} [ATLAS Collaboration],
  %``Search for single top-quark production via flavour-changing neutral currents at 8 TeV with the ATLAS detector,''
  Eur.\ Phys.\ J.\ C {\bf 76}, no. 2, 55 (2016)
  doi:10.1140/epjc/s10052-016-3876-4
  [arXiv:1509.00294 [hep-ex]].
  %%CITATION = doi:10.1140/epjc/s10052-016-3876-4;%%
  %60 citations counted in INSPIRE as of 04 May 2019


%\cite{TheATLAScollaboration:2013vha}
\bibitem{TheATLAScollaboration:2013vha} 
  The ATLAS collaboration [ATLAS Collaboration],
  %``Search for single top-quark production via FCNC in strong interaction in $\sqrt{s}=8\,\,\mathrm{TeV}$ ATLAS data,''
  ATLAS-CONF-2013-063.
  %%CITATION = ATLAS-CONF-2013-063;%%
  %34 citations counted in INSPIRE as of 02 May 2019


%\cite{Aad:2012gd}
\bibitem{Aad:2012gd} 
  G.~Aad {\it et al.} [ATLAS Collaboration],
  %``Search for FCNC single top-quark production at $\sqrt{s}=7$ TeV with the ATLAS detector,''
  Phys.\ Lett.\ B {\bf 712}, 351 (2012)
  doi:10.1016/j.physletb.2012.05.022
  [arXiv:1203.0529 [hep-ex]].
  %%CITATION = doi:10.1016/j.physletb.2012.05.022;%%
  %87 citations counted in INSPIRE as of 02 May 2019



%\cite{Abazov:2010qk}
\bibitem{Abazov:2010qk} 
  V.~M.~Abazov {\it et al.} [D0 Collaboration],
  %``Search for Flavor Changing Neutral Currents via Quark-Gluon Couplings in Single Top Quark Production using 2.3 fb$^{-1}$ of $p\bar{p}$ Collisions,''
  Phys.\ Lett.\ B {\bf 693}, 81 (2010)
  doi:10.1016/j.physletb.2010.08.011
  [arXiv:1006.3575 [hep-ex]].
  %%CITATION = doi:10.1016/j.physletb.2010.08.011;%%
  %77 citations counted in INSPIRE as of 02 May 2019



%\cite{Abazov:2007ev}
\bibitem{Abazov:2007ev} 
  V.~M.~Abazov {\it et al.} [D0 Collaboration],
  %``Search for production of single top quarks via tcg and tug flavor- changing neutral current couplings,''
  Phys.\ Rev.\ Lett.\  {\bf 99}, 191802 (2007)
  doi:10.1103/PhysRevLett.99.191802
  [hep-ex/0702005 [HEP-EX], arXiv:0801.2556 [HEP-EX]].
  %%CITATION = doi:10.1103/PhysRevLett.99.191802;%%
  %72 citations counted in INSPIRE as of 02 May 2019



%\cite{Khachatryan:2016sib}
\bibitem{Khachatryan:2016sib} 
  V.~Khachatryan {\it et al.} [CMS Collaboration],
  %``Search for anomalous Wtb couplings and flavour-changing neutral currents in t-channel single top quark production in pp collisions at $\sqrt{s} =$ 7 and 8 TeV,''
  JHEP {\bf 1702}, 028 (2017)
  doi:10.1007/JHEP02(2017)028
  [arXiv:1610.03545 [hep-ex]].
  %%CITATION = doi:10.1007/JHEP02(2017)028;%%
  %44 citations counted in INSPIRE as of 04 May 2019


%\cite{Khachatryan:2015att}
\bibitem{Khachatryan:2015att} 
  V.~Khachatryan {\it et al.} [CMS Collaboration],
  %``Search for Anomalous Single Top Quark Production in Association with a Photon in $pp$ Collisions at $ \sqrt{s}=8 $ TeV,''
  JHEP {\bf 1604}, 035 (2016)
  doi:10.1007/JHEP04(2016)035
  [arXiv:1511.03951 [hep-ex]].
  %%CITATION = doi:10.1007/JHEP04(2016)035;%%
  %63 citations counted in INSPIRE as of 04 May 2019


  %\cite{Sirunyan:2017kkr}
\bibitem{Sirunyan:2017kkr} 
  A.~M.~Sirunyan {\it et al.} [CMS Collaboration],
  %``Search for associated production of a Z boson with a single top quark and for tZ flavour-changing interactions in pp collisions at $ \sqrt{s}=8 $ TeV,''
  JHEP {\bf 1707}, 003 (2017)
  doi:10.1007/JHEP07(2017)003
  [arXiv:1702.01404 [hep-ex]].
  %%CITATION = doi:10.1007/JHEP07(2017)003;%%
  %49 citations counted in INSPIRE as of 04 May 2019



  %\cite{Aleph:2001dzz}
\bibitem{Aleph:2001dzz} 
  D.~Aleph, L3, Opal Collaborations, and the LEP Exotica Working Group,
  %``Search for single top production via flavour changing neutral currents : preliminary combined results of the LEP experiments,''
  DELPHI-2001-119 CONF 542.
  %%CITATION = DELPHI-2001-119 CONF 542;%%


%\cite{Abbiendi:2001wk}
\bibitem{Abbiendi:2001wk} 
  G.~Abbiendi {\it et al.} [OPAL Collaboration],
  %``Search for single top quark production at LEP-2,''
  Phys.\ Lett.\ B {\bf 521}, 181 (2001)
  doi:10.1016/S0370-2693(01)01195-9
  [hep-ex/0110009].
  %%CITATION = doi:10.1016/S0370-2693(01)01195-9;%%
  %86 citations counted in INSPIRE as of 02 May 2019



%\cite{Heister:2002xv}
\bibitem{Heister:2002xv} 
  A.~Heister {\it et al.} [ALEPH Collaboration],
  %``Search for single top production in $e^{+} e^{-}$ collisions at $\sqrt{s}$ up to 209-GeV,''
  Phys.\ Lett.\ B {\bf 543}, 173 (2002)
  doi:10.1016/S0370-2693(02)02307-9
  [hep-ex/0206070].
  %%CITATION = doi:10.1016/S0370-2693(02)02307-9;%%
  %67 citations counted in INSPIRE as of 02 May 2019



%\cite{Achard:2002vv}
\bibitem{Achard:2002vv} 
  P.~Achard {\it et al.} [L3 Collaboration],
  %``Search for single top production at LEP,''
  Phys.\ Lett.\ B {\bf 549}, 290 (2002)
  doi:10.1016/S0370-2693(02)02933-7
  [hep-ex/0210041].
  %%CITATION = doi:10.1016/S0370-2693(02)02933-7;%%
  %106 citations counted in INSPIRE as of 02 May 2019


%\cite{Abdallah:2003wf}
\bibitem{Abdallah:2003wf} 
  J.~Abdallah {\it et al.} [DELPHI Collaboration],
  %``Search for single top production via FCNC at LEP at $\sqrt{s}$ = 189-GeV to 208-GeV,''
  Phys.\ Lett.\ B {\bf 590}, 21 (2004)
  doi:10.1016/j.physletb.2004.03.051
  [hep-ex/0404014].
  %%CITATION = doi:10.1016/j.physletb.2004.03.051;%%
  %62 citations counted in INSPIRE as of 02 May 2019


%\cite{DELPHI:2011ab}
\bibitem{DELPHI:2011ab} 
  J.~Abdallah {\it et al.} [DELPHI Collaboration],
  %``Search for single top quark production via contact interactions at LEP2,''
  Eur.\ Phys.\ J.\ C {\bf 71}, 1555 (2011)
  doi:10.1140/epjc/s10052-011-1555-z
  [arXiv:1102.4455 [hep-ex]].
  %%CITATION = doi:10.1140/epjc/s10052-011-1555-z;%%
  %3 citations counted in INSPIRE as of 04 May 2019


%\cite{Abramowicz:2011tv}
\bibitem{Abramowicz:2011tv} 
  H.~Abramowicz {\it et al.} [ZEUS Collaboration],
  %``Search for single-top production in $ep$ collisions at HERA,''
  Phys.\ Lett.\ B {\bf 708}, 27 (2012)
  doi:10.1016/j.physletb.2012.01.025
  [arXiv:1111.3901 [hep-ex]].
  %%CITATION = doi:10.1016/j.physletb.2012.01.025;%%
  %46 citations counted in INSPIRE as of 02 May 2019



%\cite{Chekanov:2003yt}
\bibitem{Chekanov:2003yt} 
  S.~Chekanov {\it et al.} [ZEUS Collaboration],
  %``Search for single top production in ep collisions at HERA,''
  Phys.\ Lett.\ B {\bf 559}, 153 (2003)
  doi:10.1016/S0370-2693(03)00333-2
  [hep-ex/0302010].
  %%CITATION = doi:10.1016/S0370-2693(03)00333-2;%%
  %148 citations counted in INSPIRE as of 02 May 2019



%\cite{H1}
\bibitem{H1} 
  H1 Collaboration,
  ``Search for single top production in $e^{\pm}p$ collisions at HERA,''
  H1prelim-01-063 (2001).


%\cite{Aaron:2009vv}
\bibitem{Aaron:2009vv} 
  F.~D.~Aaron {\it et al.} [H1 Collaboration],
  %``Search for Single Top Quark Production at HERA,''
  Phys.\ Lett.\ B {\bf 678}, 450 (2009)
  doi:10.1016/j.physletb.2009.06.057
  [arXiv:0904.3876 [hep-ex]].
  %%CITATION = doi:10.1016/j.physletb.2009.06.057;%%
  %57 citations counted in INSPIRE as of 02 May 2019



%\cite{Aktas:2003yd}
\bibitem{Aktas:2003yd} 
  A.~Aktas {\it et al.} [H1 Collaboration],
  %``Search for single top quark production in ep collisions at HERA,''
  Eur.\ Phys.\ J.\ C {\bf 33}, 9 (2004)
  doi:10.1140/epjc/s2003-01588-2
  [hep-ex/0310032].
  %%CITATION = doi:10.1140/epjc/s2003-01588-2;%%
  %80 citations counted in INSPIRE as of 02 May 2019

%\cite{Durieux:2014xla}
\bibitem{Durieux:2014xla} 
  G.~Durieux, F.~Maltoni and C.~Zhang,
  %``Global approach to top-quark flavor-changing interactions,''
  Phys.\ Rev.\ D {\bf 91}, no. 7, 074017 (2015)
  doi:10.1103/PhysRevD.91.074017
  [arXiv:1412.7166 [hep-ph]].
  %%CITATION = doi:10.1103/PhysRevD.91.074017;%%
  %73 citations counted in INSPIRE as of 02 May 2019


  %\cite{Iltan:2002re}
\bibitem{Iltan:2002re} 
  E.~O.~Iltan and I.~Turan,
  %``The Flavor changing t ---> c l(1)- l(2)+ decay in the general two Higgs doublet model,''
  Phys.\ Rev.\ D {\bf 67}, 015004 (2003)
  doi:10.1103/PhysRevD.67.015004
  [hep-ph/0207087].
  %%CITATION = doi:10.1103/PhysRevD.67.015004;%%
  %30 citations counted in INSPIRE as of 04 May 2019

  %\cite{Frank:2006ku}
\bibitem{Frank:2006ku} 
  M.~Frank and I.~Turan,
  %``Rare decay of the top t ---> c l anti-l and single top production at ILC,''
  Phys.\ Rev.\ D {\bf 74}, 073014 (2006)
  doi:10.1103/PhysRevD.74.073014
  [hep-ph/0609069].
  %%CITATION = doi:10.1103/PhysRevD.74.073014;%%
  %37 citations counted in INSPIRE as of 04 May 2019

  %\cite{Han:2011xd}
\bibitem{Han:2011xd} 
  J.~Han, B.~Li and X.~Wang,
  %``Top quark rare three-body decays in the littlest Higgs model with T-parity,''
  Phys.\ Rev.\ D {\bf 83}, 034032 (2011)
  doi:10.1103/PhysRevD.83.034032
  [arXiv:1102.4402 [hep-ph]].
  %%CITATION = doi:10.1103/PhysRevD.83.034032;%%
  %9 citations counted in INSPIRE as of 04 May 2019


%\cite{Cerri:2018ypt}
\bibitem{Cerri:2018ypt} 
  A.~Cerri {\it et al.},
  %``Opportunities in Flavour Physics at the HL-LHC and HE-LHC,''
  arXiv:1812.07638 [hep-ph].
  %%CITATION = ARXIV:1812.07638;%%
  %20 citations counted in INSPIRE as of 02 May 2019

%\cite{AguilarSaavedra:2001ab}
\bibitem{AguilarSaavedra:2001ab} 
  J.~A.~Aguilar-Saavedra and T.~Riemann,
  %``Probing top flavor changing neutral couplings at TESLA,''
  hep-ph/0102197.
  %%CITATION = HEP-PH/0102197;%%
  %32 citations counted in INSPIRE as of 04 May 2019

%\cite{Khanpour:2014xla}
\bibitem{Khanpour:2014xla} 
  H.~Khanpour, S.~Khatibi, M.~Khatiri Yanehsari and M.~Mohammadi Najafabadi,
  %``Single top quark production as a probe of anomalous $tq\gamma$ and $tqZ$ couplings at the FCC-ee,''
  Phys.\ Lett.\ B {\bf 775}, 25 (2017)
  doi:10.1016/j.physletb.2017.10.047
  [arXiv:1408.2090 [hep-ph]].
  %%CITATION = doi:10.1016/j.physletb.2017.10.047;%%
  %27 citations counted in INSPIRE as of 04 May 2019

  %\cite{deBlas:2018mhx}
\bibitem{deBlas:2018mhx} 
  J.~de Blas {\it et al.},
  %``The CLIC Potential for New Physics,''
  CERN Yellow Rep. Monogr. Vol. 3 (2018)
  doi:10.23731/CYRM-2018-003
  [arXiv:1812.02093 [hep-ph]].
  %%CITATION = doi:10.23731/CYRM-2018-003;%%
  %15 citations counted in INSPIRE as of 04 May 2019



%\cite{Eilam:1990zc}
\bibitem{Eilam:1990zc} 
  G.~Eilam, J.~L.~Hewett and A.~Soni,
  %``Rare decays of the top quark in the standard and two Higgs doublet models,''
  Phys.\ Rev.\ D {\bf 44}, 1473 (1991)
  Erratum: [Phys.\ Rev.\ D {\bf 59}, 039901 (1999)].
  doi:10.1103/PhysRevD.44.1473, 10.1103/PhysRevD.59.039901
  %%CITATION = doi:10.1103/PhysRevD.44.1473, 10.1103/PhysRevD.59.039901;%%
  %384 citations counted in INSPIRE as of 02 May 2019

%\cite{Mele:1998ag}
\bibitem{Mele:1998ag} 
  B.~Mele, S.~Petrarca and A.~Soddu,
  %``A New evaluation of the t ---> cH decay width in the standard model,''
  Phys.\ Lett.\ B {\bf 435}, 401 (1998)
  doi:10.1016/S0370-2693(98)00822-3
  [hep-ph/9805498].
  %%CITATION = doi:10.1016/S0370-2693(98)00822-3;%%
  %197 citations counted in INSPIRE as of 02 May 2019

%\cite{AguilarSaavedra:2002ns}
\bibitem{AguilarSaavedra:2002ns} 
  J.~A.~Aguilar-Saavedra and B.~M.~Nobre,
  %``Rare top decays t ---> c gamma, t ---> cg and CKM unitarity,''
  Phys.\ Lett.\ B {\bf 553}, 251 (2003)
  doi:10.1016/S0370-2693(02)03230-6
  [hep-ph/0210360].
  %%CITATION = doi:10.1016/S0370-2693(02)03230-6;%%
  %101 citations counted in INSPIRE as of 02 May 2019

%decay Tevatron


%\cite{AguilarSaavedra:2004wm}
\bibitem{AguilarSaavedra:2004wm} 
  J.~A.~Aguilar-Saavedra,
  %``Top flavor-changing neutral interactions: Theoretical expectations and experimental detection,''
  Acta Phys.\ Polon.\ B {\bf 35}, 2695 (2004)
  [hep-ph/0409342].
  %%CITATION = HEP-PH/0409342;%%
  %311 citations counted in INSPIRE as of 02 May 2019

%\cite{Weinberg:1978kz}
\bibitem{Weinberg:1978kz} 
  S.~Weinberg,
  %``Phenomenological Lagrangians,''
  Physica A {\bf 96}, no. 1-2, 327 (1979).
  doi:10.1016/0378-4371(79)90223-1
  %%CITATION = doi:10.1016/0378-4371(79)90223-1;%%
  %3233 citations counted in INSPIRE as of 21 May 2019

  %\cite{Leung:1984ni}
\bibitem{Leung:1984ni} 
  C.~N.~Leung, S.~T.~Love and S.~Rao,
  %``Low-Energy Manifestations of a New Interaction Scale: Operator Analysis,''
  Z.\ Phys.\ C {\bf 31}, 433 (1986).
  doi:10.1007/BF01588041
  %%CITATION = doi:10.1007/BF01588041;%%
  %310 citations counted in INSPIRE as of 21 May 2019

  %\cite{Buchmuller:1985jz}
\bibitem{Buchmuller:1985jz} 
  W.~Buchmuller and D.~Wyler,
  %``Effective Lagrangian Analysis of New Interactions and Flavor Conservation,''
  Nucl.\ Phys.\ B {\bf 268}, 621 (1986).
  doi:10.1016/0550-3213(86)90262-2
  %%CITATION = doi:10.1016/0550-3213(86)90262-2;%%
  %1535 citations counted in INSPIRE as of 21 May 2019



%\cite{AguilarSaavedra:2018nen}
\bibitem{AguilarSaavedra:2018nen} 
  J.~A.~Aguilar-Saavedra {\it et al.},
  %``Interpreting top-quark LHC measurements in the standard-model effective field theory,''
  arXiv:1802.07237 [hep-ph].
  %%CITATION = ARXIV:1802.07237;%%
  %37 citations counted in INSPIRE as of 02 May 2019
 
%\cite{Grzadkowski:2010es}
\bibitem{Grzadkowski:2010es} 
  B.~Grzadkowski, M.~Iskrzynski, M.~Misiak and J.~Rosiek,
  %``Dimension-Six Terms in the Standard Model Lagrangian,''
  JHEP {\bf 1010}, 085 (2010)
  doi:10.1007/JHEP10(2010)085
  [arXiv:1008.4884 [hep-ph]].
  %%CITATION = doi:10.1007/JHEP10(2010)085;%%
  %869 citations counted in INSPIRE as of 02 May 2019

%\cite{BarShalom:1999iy}
\bibitem{BarShalom:1999iy} 
  S.~Bar-Shalom and J.~Wudka,
  %``Flavor changing single top quark production channels at e+ e- colliders in the effective Lagrangian description,''
  Phys.\ Rev.\ D {\bf 60}, 094016 (1999)
  doi:10.1103/PhysRevD.60.094016
  [hep-ph/9905407].
  %%CITATION = doi:10.1103/PhysRevD.60.094016;%%
  %44 citations counted in INSPIRE as of 02 May 2019

%\cite{Chala:2018agk}
\bibitem{Chala:2018agk} 
  M.~Chala, J.~Santiago and M.~Spannowsky,
  %``Constraining four-fermion operators using rare top decays,''
  JHEP {\bf 1904}, 014 (2019)
  doi:10.1007/JHEP04(2019)014
  [arXiv:1809.09624 [hep-ph]].
  %%CITATION = doi:10.1007/JHEP04(2019)014;%%
  %6 citations counted in INSPIRE as of 04 May 2019

%\cite{Davidson:2015zza}
\bibitem{Davidson:2015zza} 
  S.~Davidson, M.~L.~Mangano, S.~Perries and V.~Sordini,
  %``Lepton Flavour Violating top decays at the LHC,''
  Eur.\ Phys.\ J.\ C {\bf 75}, no. 9, 450 (2015)
  doi:10.1140/epjc/s10052-015-3649-5
  [arXiv:1507.07163 [hep-ph]].
  %%CITATION = doi:10.1140/epjc/s10052-015-3649-5;%%
  %15 citations counted in INSPIRE as of 04 May 2019

  %\cite{ATLAS:2018avw}
\bibitem{ATLAS:2018avw} 
  The ATLAS collaboration [ATLAS Collaboration],
  %``Search for charged lepton-flavour violation in top-quark decays at the LHC with the ATLAS detector,''
  ATLAS-CONF-2018-044.
  %%CITATION = ATLAS-CONF-2018-044;%%
  %3 citations counted in INSPIRE as of 04 May 2019

%\cite{Alcaide:2019pnf}
\bibitem{Alcaide:2019pnf} 
  J.~Alcaide, S.~Banerjee, M.~Chala and A.~Titov,
  %``Probes of the Standard Model effective field theory extended with a right-handed neutrino,''
  arXiv:1905.11375 [hep-ph].
  %%CITATION = ARXIV:1905.11375;%%


%\cite{Alwall:2014hca}
\bibitem{Alwall:2014hca} 
  J.~Alwall {\it et al.},
  %``The automated computation of tree-level and next-to-leading order differential cross sections, and their matching to parton shower simulations,''
  JHEP {\bf 1407}, 079 (2014)
  doi:10.1007/JHEP07(2014)079
  [arXiv:1405.0301 [hep-ph]].
  %%CITATION = doi:10.1007/JHEP07(2014)079;%%
  %3577 citations counted in INSPIRE as of 04 May 2019

  %\cite{Sjostrand:2006za}
\bibitem{Sjostrand:2006za} 
  T.~Sjostrand, S.~Mrenna and P.~Z.~Skands,
  %``PYTHIA 6.4 Physics and Manual,''
  JHEP {\bf 0605}, 026 (2006)
  doi:10.1088/1126-6708/2006/05/026
  [hep-ph/0603175].
  %%CITATION = doi:10.1088/1126-6708/2006/05/026;%%
  %10047 citations counted in INSPIRE as of 04 May 2019
%\cite{Sjostrand:2007gs}
\bibitem{Sjostrand:2007gs} 
  T.~Sjostrand, S.~Mrenna and P.~Z.~Skands,
  %``A Brief Introduction to PYTHIA 8.1,''
  Comput.\ Phys.\ Commun.\  {\bf 178}, 852 (2008)
  doi:10.1016/j.cpc.2008.01.036
  [arXiv:0710.3820 [hep-ph]].
  %%CITATION = doi:10.1016/j.cpc.2008.01.036;%%
  %4350 citations counted in INSPIRE as of 04 May 2019

  %\cite{Alloul:2013bka}
\bibitem{Alloul:2013bka} 
  A.~Alloul, N.~D.~Christensen, C.~Degrande, C.~Duhr and B.~Fuks,
  %``FeynRules  2.0 - A complete toolbox for tree-level phenomenology,''
  Comput.\ Phys.\ Commun.\  {\bf 185}, 2250 (2014)
  doi:10.1016/j.cpc.2014.04.012
  [arXiv:1310.1921 [hep-ph]].
  %%CITATION = doi:10.1016/j.cpc.2014.04.012;%%
  %1125 citations counted in INSPIRE as of 24 May 2019
  %\cite{Degrande:2011ua}
\bibitem{Degrande:2011ua} 
  C.~Degrande, C.~Duhr, B.~Fuks, D.~Grellscheid, O.~Mattelaer and T.~Reiter,
  %``UFO - The Universal FeynRules Output,''
  Comput.\ Phys.\ Commun.\  {\bf 183}, 1201 (2012)
  doi:10.1016/j.cpc.2012.01.022
  [arXiv:1108.2040 [hep-ph]].
  %%CITATION = doi:10.1016/j.cpc.2012.01.022;%%
  %627 citations counted in INSPIRE as of 24 May 2019

  %\cite{deFavereau:2013fsa}
\bibitem{deFavereau:2013fsa} 
  J.~de Favereau {\it et al.} [DELPHES 3 Collaboration],
  %``DELPHES 3, A modular framework for fast simulation of a generic collider experiment,''
  JHEP {\bf 1402}, 057 (2014)
  doi:10.1007/JHEP02(2014)057
  [arXiv:1307.6346 [hep-ex]].
  %%CITATION = doi:10.1007/JHEP02(2014)057;%%
  %1217 citations counted in INSPIRE as of 04 May 2019

%\cite{Cacciari:2011ma}
\bibitem{Cacciari:2011ma} 
  M.~Cacciari, G.~P.~Salam and G.~Soyez,
  %``FastJet User Manual,''
  Eur.\ Phys.\ J.\ C {\bf 72}, 1896 (2012)
  doi:10.1140/epjc/s10052-012-1896-2
  [arXiv:1111.6097 [hep-ph]].
  %%CITATION = doi:10.1140/epjc/s10052-012-1896-2;%%
  %2851 citations counted in INSPIRE as of 21 May 2019

  %\cite{Cacciari:2008gp}
\bibitem{Cacciari:2008gp} 
  M.~Cacciari, G.~P.~Salam and G.~Soyez,
  %``The anti-$k_t$ jet clustering algorithm,''
  JHEP {\bf 0804}, 063 (2008)
  doi:10.1088/1126-6708/2008/04/063
  [arXiv:0802.1189 [hep-ph]].
  %%CITATION = doi:10.1088/1126-6708/2008/04/063;%%
  %5884 citations counted in INSPIRE as of 21 May 2019

%\cite{Degrande:2014tta}
\bibitem{Degrande:2014tta} 
  C.~Degrande, F.~Maltoni, J.~Wang and C.~Zhang,
  %``Automatic computations at next-to-leading order in QCD for top-quark flavor-changing neutral processes,''
  Phys.\ Rev.\ D {\bf 91}, 034024 (2015)
  doi:10.1103/PhysRevD.91.034024
  [arXiv:1412.5594 [hep-ph]].
  %%CITATION = doi:10.1103/PhysRevD.91.034024;%%
  %42 citations counted in INSPIRE as of 04 May 2019

%\cite{Liu:2005dp}
\bibitem{Liu:2005dp} 
  J.~J.~Liu, C.~S.~Li, L.~L.~Yang and L.~G.~Jin,
  %``Next-to-leading order QCD corrections to the direct top quark production via model-independent FCNC couplings at hadron colliders,''
  Phys.\ Rev.\ D {\bf 72}, 074018 (2005)
  doi:10.1103/PhysRevD.72.074018
  [hep-ph/0508016].
  %%CITATION = doi:10.1103/PhysRevD.72.074018;%%
  %50 citations counted in INSPIRE as of 23 May 2019

  %\cite{Gao:2009rf}
\bibitem{Gao:2009rf} 
  J.~Gao, C.~S.~Li, J.~J.~Zhang and H.~X.~Zhu,
  %``Next-to-leading order QCD corrections to the single top quark production via model-independent t-q-g flavor-changing neutral-current couplings at hadron colliders,''
  Phys.\ Rev.\ D {\bf 80}, 114017 (2009)
  doi:10.1103/PhysRevD.80.114017
  [arXiv:0910.4349 [hep-ph]].
  %%CITATION = doi:10.1103/PhysRevD.80.114017;%%
  %23 citations counted in INSPIRE as of 23 May 2019

%\cite{Zhang:2011gh}
\bibitem{Zhang:2011gh} 
  Y.~Zhang, B.~H.~Li, C.~S.~Li, J.~Gao and H.~X.~Zhu,
  %``Next-to-leading order QCD corrections to the top quark associated with $\gamma$ production via model-independent flavor-changing neutral-current couplings at hadron colliders,''
  Phys.\ Rev.\ D {\bf 83}, 094003 (2011)
  doi:10.1103/PhysRevD.83.094003
  [arXiv:1101.5346 [hep-ph]].
  %%CITATION = doi:10.1103/PhysRevD.83.094003;%%
  %21 citations counted in INSPIRE as of 23 May 2019

%\cite{Li:2011ek}
\bibitem{Li:2011ek} 
  B.~H.~Li, Y.~Zhang, C.~S.~Li, J.~Gao and H.~X.~Zhu,
  %``Next-to-leading order QCD corrections to $tZ$ associated production via the flavor-changing neutral-current couplings at hadron colliders,''
  Phys.\ Rev.\ D {\bf 83}, 114049 (2011)
  doi:10.1103/PhysRevD.83.114049
  [arXiv:1103.5122 [hep-ph]].
  %%CITATION = doi:10.1103/PhysRevD.83.114049;%%
  %18 citations counted in INSPIRE as of 23 May 2019

  %\cite{Wang:2012gp}
\bibitem{Wang:2012gp} 
  Y.~Wang, F.~P.~Huang, C.~S.~Li, B.~H.~Li, D.~Y.~Shao and J.~Wang,
  %``Constraints on flavor-changing neutral-current $Htq$ couplings from the signal of $tH$ associated production with QCD next-to-leading order accuracy at the LHC,''
  Phys.\ Rev.\ D {\bf 86}, 094014 (2012)
  doi:10.1103/PhysRevD.86.094014
  [arXiv:1208.2902 [hep-ph]].
  %%CITATION = doi:10.1103/PhysRevD.86.094014;%%
  %26 citations counted in INSPIRE as of 23 May 2019

%\cite{Drobnak:2010wh}
\bibitem{Drobnak:2010wh} 
  J.~Drobnak, S.~Fajfer and J.~F.~Kamenik,
  %``Flavor Changing Neutral Coupling Mediated Radiative Top Quark Decays at Next-to-Leading Order in QCD,''
  Phys.\ Rev.\ Lett.\  {\bf 104}, 252001 (2010)
  doi:10.1103/PhysRevLett.104.252001
  [arXiv:1004.0620 [hep-ph]].
  %%CITATION = doi:10.1103/PhysRevLett.104.252001;%%
  %41 citations counted in INSPIRE as of 23 May 2019

  %\cite{Drobnak:2010by}
\bibitem{Drobnak:2010by} 
  J.~Drobnak, S.~Fajfer and J.~F.~Kamenik,
  %``QCD Corrections to Flavor Changing Neutral Coupling Mediated Rare Top Quark Decays,''
  Phys.\ Rev.\ D {\bf 82}, 073016 (2010)
  doi:10.1103/PhysRevD.82.073016
  [arXiv:1007.2551 [hep-ph]].
  %%CITATION = doi:10.1103/PhysRevD.82.073016;%%
  %26 citations counted in INSPIRE as of 23 May 2019

  %\cite{Zhang:2013xya}
\bibitem{Zhang:2013xya} 
  C.~Zhang and F.~Maltoni,
  %``Top-quark decay into Higgs boson and a light quark at next-to-leading order in QCD,''
  Phys.\ Rev.\ D {\bf 88}, 054005 (2013)
  doi:10.1103/PhysRevD.88.054005
  [arXiv:1305.7386 [hep-ph]].
  %%CITATION = doi:10.1103/PhysRevD.88.054005;%%
  %58 citations counted in INSPIRE as of 23 May 2019

  %\cite{Zhang:2014rja}
\bibitem{Zhang:2014rja} 
  C.~Zhang,
  %``Effective field theory approach to top-quark decay at next-to-leading order in QCD,''
  Phys.\ Rev.\ D {\bf 90}, no. 1, 014008 (2014)
  doi:10.1103/PhysRevD.90.014008
  [arXiv:1404.1264 [hep-ph]].
  %%CITATION = doi:10.1103/PhysRevD.90.014008;%%
  %40 citations counted in INSPIRE as of 23 May 2019

  %\cite{Ruan:2018yrh}
\bibitem{Ruan:2018yrh} 
  M.~Ruan {\it et al.},
  %``Reconstruction of physics objects at the Circular Electron Positron Collider with Arbor,''
  Eur.\ Phys.\ J.\ C {\bf 78}, no. 5, 426 (2018)
  doi:10.1140/epjc/s10052-018-5876-z
  [arXiv:1806.04879 [hep-ex]].
  %%CITATION = doi:10.1140/epjc/s10052-018-5876-z;%%
  %6 citations counted in INSPIRE as of 01 Jun 2019


  %\cite{Atwood:1991ka}
\bibitem{Atwood:1991ka} 
  D.~Atwood and A.~Soni,
  %``Analysis for magnetic moment and electric dipole moment form-factors of the top quark via e+ e- ---> t anti-t,''
  Phys.\ Rev.\ D {\bf 45}, 2405 (1992).
  doi:10.1103/PhysRevD.45.2405
  %%CITATION = doi:10.1103/PhysRevD.45.2405;%%
  %234 citations counted in INSPIRE as of 04 May 2019

  %\cite{Diehl:1993br}
\bibitem{Diehl:1993br} 
  M.~Diehl and O.~Nachtmann,
  %``Optimal observables for the measurement of three gauge boson couplings in e+ e- ---> W+ W-,''
  Z.\ Phys.\ C {\bf 62}, 397 (1994).
  doi:10.1007/BF01555899
  %%CITATION = doi:10.1007/BF01555899;%%
  %174 citations counted in INSPIRE as of 04 May 2019

%\cite{Liu:2019wmi}
\bibitem{Liu:2019wmi} 
  W.~Liu and H.~Sun,
  %``Top FCNC interactions through dimension six four-fermion operators at the electron proton collider,''
  arXiv:1906.04884 [hep-ph].
  %%CITATION = ARXIV:1906.04884;%%

\end{thebibliography}
\end{document}